%% file: main.tex
\documentclass[11pt,letterpaper]{article}
\usepackage[margin=1in]{geometry}
\usepackage[T1]{fontenc}
\usepackage[utf8]{inputenc}
\input{math_commands.tex}

\usepackage{amssymb,amsthm}

\usepackage{booktabs,graphicx,float,microtype,multirow,tabularx,xcolor}
\usepackage{longtable,array,verbatim,subcaption}
\usepackage[authoryear,round]{natbib}
\setcitestyle{citesep={;},aysep={,},yysep={;}}
\usepackage{xurl}
\usepackage{hyperref}

\hypersetup{
  hidelinks,
  pdfauthor={Wenlong Yu, Bingxu Lu, Qilong Wang, Qinghua Hu, Qianru Sun},
  pdftitle={From Local Errors to System Failures: Modeling and Benchmarking Error Propagation in Multi-Agent LLM Systems},
  pdfsubject={Research preprint}
}
\graphicspath{{figures/}}

\title{From Local Errors to System Failures:\\
Modeling and Benchmarking Error Propagation in\\
Multi-Agent LLM Systems}
\author{
  {\normalsize Wenlong Yu$^{1}$ \quad
  Bingxu Lu$^{2}$\thanks{Corresponding author. Emails: \href{mailto:wloongyu@gmail.com}{\texttt{wloongyu@gmail.com}}, \href{mailto:lubingxu@nankai.edu.cn}{\texttt{lubingxu@nankai.edu.cn}}.} \quad
  Qilong Wang$^{3}$ \quad Qinghua Hu$^{3}$ \quad Qianru Sun$^{1}$} \\[0.5em]
  {\scriptsize $^{1}$Singapore Management University, Singapore \quad
  $^{2}$Nankai University, Tianjin, China \quad
  $^{3}$Tianjin University, Tianjin, China}
}
\date{}

\begin{document}
\maketitle

\input{sections/00_abstract}

\input{sections/01_introduction}

\input{sections/02_related_work}

\input{sections/03_repa_formulation_analysis}

\input{sections/05_results}

\input{sections/07_conclusion}

\input{sections/08_statements}

\bibliography{references}
\bibliographystyle{plainnat}

\clearpage
\appendix
\setcounter{table}{0}
\renewcommand{\thetable}{S\arabic{table}}
\renewcommand{\theHtable}{S\arabic{table}}

\input{appendix/A_repa_formulation_analysis}
\input{appendix/B_literature_comparison}

\input{appendix/C_experimental_details}
\input{appendix/D_prompt_templates}
\input{appendix/E_limitations_future}

\end{document}

%% file: math_commands.tex
\usepackage{amsmath,amsfonts,bm}

\def\eqref#1{equation~\ref{#1}}

\def\1{\bm{1}}

\DeclareMathAlphabet{\mathsfit}{\encodingdefault}{\sfdefault}{m}{sl}
\SetMathAlphabet{\mathsfit}{bold}{\encodingdefault}{\sfdefault}{bx}{n}



%% file: sections/00_abstract.tex
\begin{abstract}

In multi-agent systems (MAS) built on large language models (LLMs), errors can propagate through intermediate messages and compromise final answers.
However, outcome-based evaluation obscures how agents amplify, attenuate, or correct these errors.
We propose Response-Aware Error Propagation among Agents (REPA), a structural model that connects measured local responses to system-level error propagation. 
REPA separately estimates input assembly and agent processing across multiple error channels, while theoretical analysis relates local modeling errors to endpoint prediction error.
Experiments span three Qwen base models, six MAS workflows, and ten knowledge and reasoning settings in ErrPropBench-10. 
REPA achieves a mean condition-level Spearman correlation of 0.84 across base models, compared with 0.47 for direct regression. 
Controlled experiments also demonstrate its utility for targeted recovery. 
REPA-guided retrieval-augmented generation (RAG) improves accuracy over no retrieval by an average of 19 percentage points with 16\% additional tokens. 
By linking local agent behavior to system susceptibility, REPA provides an interpretable foundation for studying error propagation and developing targeted optimization in MAS.

\end{abstract}

%% file: sections/01_introduction.tex
\section{Introduction}
\label{sec:introduction}

Multi-agent systems (MAS) built on large language models (LLMs) address complex tasks through role specialization and structured communication \citep{wu2023autogen,li2023camel,hong2024metagpt}.
Within these workflows, downstream agents incorporate intermediate results from upstream agents into their reasoning and decision-making.
This dependence, however, also creates pathways through which inaccurate external information or erroneous agent responses can propagate and compromise the final answer~\citep{huang2025resilience,jia2026masfire}.
Understanding MAS reliability therefore requires examining how individual agents respond to erroneous information and how workflow dependencies link these local responses to the final task outcome.


%
This local-to-system relationship cannot be captured by final-answer correctness alone~\citep{liu2026mosaic,jia2026masfire}.
An agent may reject, revise, or adopt misleading information, and the workflow then determines which downstream agents receive its response \citep{huang2025resilience,shen2025eib}.
Moreover, different patterns of error adoption and correction can produce the same final correctness label, obscuring how erroneous information was handled within the workflow.
Therefore, a propagation analysis should examine intermediate inputs and outputs alongside task outcomes, distinguishing changes in message content from changes in task performance.

Recent studies have advanced failure diagnosis and robustness evaluation in MAS \citep{cemri2025mast,zhang2025whowhen,jia2026masfire,emde2026maseval}, while others examine how communication topology and repeated interactions shape information propagation and hallucination dynamics \citep{shen2025eib,jamshidi2026hallucinationcascade,jamshidi2026collective}.
Some optimization approaches require additional training, whereas post-hoc attribution relies on completed execution traces \citep{shen2025eib,zhou2026resmas,zhang2025whowhen}.
A complementary question is whether system-level propagation can be predicted from local measurements without fine-tuning the underlying LLMs or observing downstream test outputs.
Answering this question requires distinguishing how erroneous information changes an agent's input from how the agent responds across multiple error channels.
These local responses must then be composed through downstream processing and aggregation, since local susceptibility alone does not determine the final system response.
The central challenge is therefore to connect separately estimated local responses with independently validated system-level predictions.

To connect measured local responses with system-level predictions, we propose Response-Aware Error Propagation among Agents (REPA), a structural model of error propagation across multiple error channels.
Reference and perturbed executions of the same instance are compared under matched conditions, yielding continuous-valued changes at corresponding agent inputs and outputs for separately estimating input-assembly and agent-processing responses.
REPA then composes these locally estimated response models along the executed workflow, explicitly accounting for aggregation and finite-round feedback.
Once fitted, it predicts system-level propagation from source measurements and execution structure, without fine-tuning the underlying LLMs or observing test-time intermediate states or final outputs.
Under the modeling assumptions, theoretical analysis characterizes this composition and relates local modeling errors to endpoint prediction error.

Our evaluation covers six MAS workflows and single-agent baselines using open-source foundation models of varying sizes \citep{qwen2024qwen25,qwen2026qwen35}.
To establish a reliable measurement basis for local response estimation and system-level validation, we first assess observer agreement with independent reference judgments, then evaluate REPA's predictions within fixed conditions.
To extend the evaluation across multiple knowledge and reasoning tasks, we construct ErrPropBench-10, covering seven domains and three multi-hop question-answering datasets.
Further experiments demonstrate that REPA-derived estimates can guide node selection, targeted prompt restoration, and retrieval-augmented generation (RAG), supporting both agent-level refinement and MAS-level optimization at a reasonable inference cost.
Overall, the main contributions are as follows:

\begin{itemize}
\item \textbf{A local-to-system model of error propagation.}
We propose REPA, a structural model of error propagation in MAS based on continuous-valued responses across multiple error channels.
REPA characterizes how errors at individual agents are transformed through downstream processing and aggregation.
By composing locally estimated agent responses along the executed topology, it predicts system-level susceptibility to error propagation.


\item \textbf{Local response estimation and compositional analysis.}
Motivated by Lipschitz sensitivity analysis, REPA estimates input-assembly and agent-processing responses from paired observations.
Theoretical analysis derives explicit expressions for topology-dependent error propagation, showing how propagation predictions depend on perturbation placement and output selection, in addition to the communication topology.

\item \textbf{Guided recovery, enhancement, and a controlled benchmark.}
Recovery studies show that prompt refinement and RAG at candidate nodes identified by REPA improve MAS performance at modest inference cost.
We also develop a benchmark for paired evaluation of error propagation across ten knowledge and reasoning settings.

\end{itemize}

%% file: sections/02_related_work.tex
\section{Related Work}
\label{sec:related}

\paragraph{Failure diagnosis and propagation modeling.}
MAST characterizes failure modes~\citep{cemri2025mast}, while Who\&When attributes failures from execution traces~\citep{zhang2025whowhen}. Controlled-fault evaluations assess robustness~\citep{huang2025resilience,jia2026masfire,yu2025netsafe}, and MASEval evaluates system-level performance~\citep{emde2026maseval}.
Structural studies examine influence-driven topology learning, uncertainty preservation, hallucination dynamics, and preference transmission~\citep{shen2025eib,emde2026lost,jamshidi2026hallucinationcascade,jamshidi2026collective,liu2026contagion,niu2026reliability}.
These studies establish the importance of interaction structure, although their influence, uncertainty, and preference measures target different quantities from paired semantic-error changes.

\paragraph{Agent refinement and targeted recovery.}
Self-Refine and Reflexion use feedback to revise outputs, while RAG and Self-RAG supply retrieved evidence~\citep{madaan2023selfrefine,shinn2023reflexion,lewis2020retrieval,asai2023selfrag}.
ResMAS combines learned topology optimization with prompt refinement; MASC detects anomalies through next-execution reconstruction and triggers correction~\citep{zhou2026resmas,shen2026masc}.
Self-Refine and Reflexion operate through additional inference rather than parameter updates, whereas the full ResMAS and MASC pipelines include learned components.
These approaches improve task outcomes through refinement or correction, but do not directly test whether locally measured responses can predict system-level error propagation.

REPA connects these research directions through a local-to-system model.
Its key distinction is the separate estimation of input-assembly and agent-processing responses across multiple error channels, composed through downstream workflows.
Once fitted, the model predicts system responses without fine-tuning the underlying LLMs or observing test-time intermediate states and final outputs.
Theoretical analysis links local modeling errors to end-to-end prediction error, while controlled restoration and evidence-placement experiments demonstrate REPA's ability to guide targeted optimization of agents and MAS.
Appendix~\ref{app:repa-literature-scope} gives more comparison details.

%% file: sections/03_repa_formulation_analysis.tex
\section{REPA: Formulation and Analysis}
\label{sec:method}

\begin{figure}[t!]
\centering
\includegraphics[width=\linewidth]{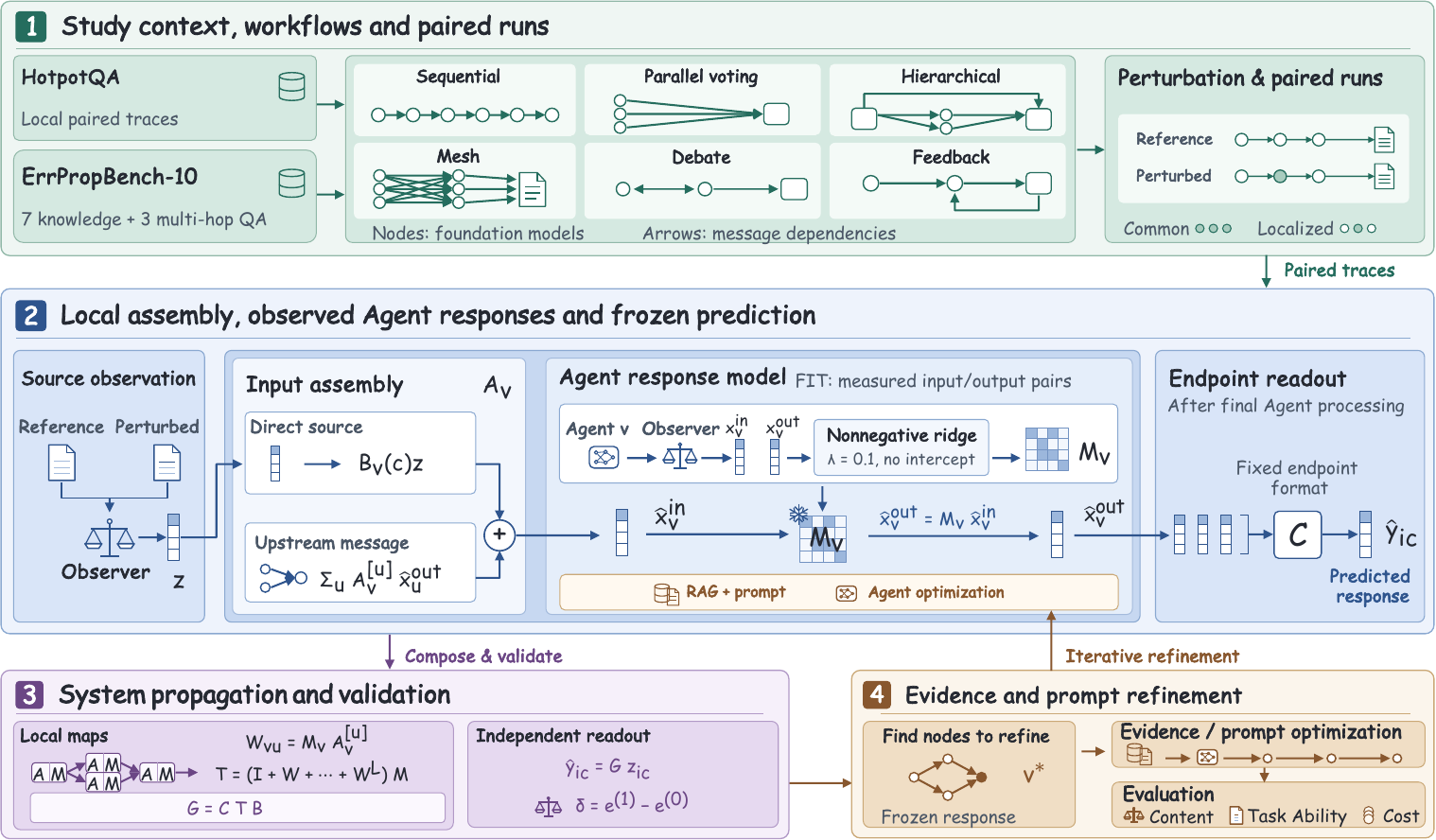}
\caption{\textbf{Overview of REPA and its evaluation workflow.}
REPA composes input-assembly and agent-response models along the executed workflow to predict system-level error responses.
The diagnostics inform prompt and evidence refinement for improved agent and MAS performance.}
\label{fig:framework}
\vspace{-10pt}
\end{figure}

As illustrated in Fig.~\ref{fig:framework}, REPA predicts system-level responses to a source perturbation by composing locally estimated input-assembly and agent-response maps.
Paired measurements distinguish changes in agent inputs from output responses.
The resulting diagnostics inform targeted refinement to improve task performance at reasonable inference cost.
Theoretical analysis establishes the finite-execution composition and relates local modeling errors to system-level prediction error. 
Further details on estimation and theoretical derivations are provided in Appendix~\ref{app:repa-theory}.

\subsection{Paired Measurement and Prediction Task}

An experimental condition $c$ specifies the model configuration, perturbation and placement, workflow, and observation and readout rules.
For each instance $i$, the reference execution $t_{icv}^{(0),s}$ and the perturbed execution $t_{icv}^{(1),s}$ use the same configuration except for the specified intervention, although the reference need not be correct.
Each agent invocation forms a separate node $v$, and finite-round feedback is unfolded in execution order to obtain a directed acyclic graph (DAG) $\mathcal G_c$.

Let $s\in\{\mathrm{in},\mathrm{out}\}$ identify the input or output at node $v$.
A fixed observer $\psi$ uses a paired scoring function $\mathcal D_\psi$ to measure content discrepancy across $d$ error channels,
\begin{equation}
\mathbf x_{icv}^{s}
=\mathcal D_\psi\!\left(t_{icv}^{(0),s},t_{icv}^{(1),s}\right)
\in[0,1]^d.
\label{eq:repa-state}
\end{equation}
Each coordinate of $\mathbf x_{icv}^{s}$ quantifies the discrepancy in one channel.
REPA predicts content discrepancies from paired observer measurements without fine-tuning the underlying LLMs.
The observer separately scores error severity in reference and perturbed outputs to distinguish increases in error severity from correction, while a task judge evaluates final-answer correctness.

The source discrepancy $\mathbf z_{ic}$ is the input discrepancy measured at the designated source node,
$\mathbf z_{ic}:=\mathbf x_{ic,v_{\mathrm{src}}(c)}^{\mathrm{in}}$.
This node is the first invocation under common exposure and the designated recipient under localized exposure.
Starting from the measured source discrepancy, REPA recursively predicts downstream discrepancy vectors, using predicted rather than measured parent-output discrepancies.
The final prediction is compared with the independently measured output discrepancy.

\subsection{Input Assembly and Local Response Estimation}

REPA separately models how source perturbations and upstream messages affect an agent's input and how the agent transforms the resulting input discrepancy into an output response.
For node $v$, the assembly vector $\mathbf f_{icv}\in\mathbb R_+^{h_v}$ collects source and parent-output discrepancies, where $h_v$ is its dimension.
Omitting $i$ and $c$ for clarity, input assembly and agent response are modeled as
\begin{equation}
\mathbf x_v^{\mathrm{in}}=A_v\mathbf f_v+\boldsymbol\epsilon_v^{\mathrm{in}},
\qquad
\mathbf x_v^{\mathrm{out}}=M_v\mathbf x_v^{\mathrm{in}}+\boldsymbol\epsilon_v^{\mathrm{out}}.
\label{eq:repa-local}
\end{equation}
The input-assembly map $A_v\in\mathbb R_+^{d\times h_v}$ transforms this vector into the $d$-channel discrepancy.
In the agent-response map $M_v\in\mathbb R_+^{d\times d}$, rows index output channels, columns index input channels, and off-diagonal entries capture cross-channel predictive associations.
The residuals account for variation not explained by assembly and processing (e.g., measurement error).
Lipschitz-type sensitivity analysis and Taylor approximation motivate the local linear representation under suitable smoothness assumptions \citep{khalil2002nonlinear, nocedal2006numerical}.
In REPA, the response maps are estimated from paired observations rather than computed as exact Jacobians or certified sensitivity bounds.

The two maps are estimated separately from observed local states on the fitting set:
\begin{equation}
\begin{aligned}
A_v&=\arg\min_{A\geq0}\sum_{(i,c)\in\mathcal F_v}
  \|\mathbf x_{icv}^{\mathrm{in}}-A\mathbf f_{icv}\|_2^2+\lambda\|A\|_F^2,\\
M_v&=\arg\min_{M\geq0}\sum_{(i,c)\in\mathcal F_v}
  \|\mathbf x_{icv}^{\mathrm{out}}-M\mathbf x_{icv}^{\mathrm{in}}\|_2^2+\lambda\|M\|_F^2.
\end{aligned}
\label{eq:repa-fit}
\end{equation}
The fitting set $\mathcal F_v$ contains paired observations across perturbations and placements for a fixed workflow.
The response maps are estimated using nonnegative ridge regression, with regularization to stabilize coefficient estimation~\citep{hoerl1970ridge}.
Once the maps are frozen, prediction proceeds in workflow order, replacing measured parent outputs in $\mathbf f_v$ with their predicted values:
\begin{equation}
\widehat{\mathbf x}_v^{\mathrm{in}}=A_v\widehat{\mathbf f}_v,
\qquad
\widehat{\mathbf x}_v^{\mathrm{out}}=M_v\widehat{\mathbf x}_v^{\mathrm{in}}.
\label{eq:repa-recursion}
\end{equation}

\subsection{Execution-Aware Composition and Readout}
\label{sec:repa-theory}

To express the node-wise predictions at the system level, stack the output-discrepancy vectors of all $N$ invocations into $\mathbf x^{\mathrm{out}}\in\mathbb R^{Nd}$.
The block-diagonal matrix $M=\operatorname{diag}(M_1,\ldots,M_N)$ collects their agent-response maps.
For the specified perturbation placement, the matrix
$B \in \mathbb{R}_{+}^{Nd \times p}$ collects the active
source-input blocks of the assembly maps $A_v$, where $p$ denotes the source dimension ($p=d$ for the source discrepancy defined above).
For each executed connection $u\to v$, let $A_v^{[u]}$ denote the block that maps node $u$'s output discrepancy into $v$'s input discrepancy.
The corresponding propagation block is $W_{vu}=M_vA_v^{[u]}$, with $W_{vu}=0$ when no such connection exists.
Thus, $MB\mathbf z$ represents the direct source contribution after agent processing, while $W$ carries parent-output discrepancies through subsequent assembly and processing.
The recursive prediction can then be written as
\begin{equation}
\widehat{\mathbf x}^{\mathrm{out}}
=MB\mathbf z+W\widehat{\mathbf x}^{\mathrm{out}},
\qquad
\widehat{\mathbf y}
=C\widehat{\mathbf x}^{\mathrm{out}}.
\label{eq:repa-stacked}
\end{equation}
The execution order makes $W$ strictly block-lower-triangular, enabling sequential prediction.
$C$ selects the final output coordinates after the corresponding agent-response models.

Unrolling the recursion maps source perturbations to final output discrepancies.
The following identity shows how this mapping combines the fitted local responses with the execution structure.
For a fixed invocation DAG whose longest directed path contains $L$ edges,
\begin{equation}
S=\sum_{k=0}^{L}W^k,\qquad
T=SM,\qquad
G=CTB,\qquad
\widehat{\mathbf y}=G\mathbf z.
\label{eq:repa-composition}
\end{equation}
Each term $W^k$ collects contributions along executed paths of length $k$.
Together with the initial processing factor $M$, these contributions form $T$, while $G=CTB$ incorporates source placement and final readout.
The system-level map $G$ is constructed by composing the fitted $A_v$ and $M_v$ along the executed workflow, without additional global fitting.
This finite-execution identity requires no infinite-horizon stability assumption and does not imply that text generation itself is linear.

To summarize system susceptibility under condition $c$, we aggregate the observed and predicted response magnitudes over the same set of eligible samples $\mathcal I_c$,
\begin{equation}
s_c=\frac{\sum_{i\in\mathcal I_c}\|\mathbf y_{ic}\|_2}{Z_c},
\qquad
\widehat s_c=\frac{\sum_{i\in\mathcal I_c}\|\widehat{\mathbf y}_{ic}\|_2}{Z_c},
\qquad
\mathbf y_{ic}=C_c\mathbf x_{ic}^{\mathrm{out}}.
\label{eq:repa-score}
\end{equation}
Here $s_c$ and $\widehat s_c$ denote measured and predicted susceptibility, respectively.
The shared denominator $Z_c>0$ normalizes for source-perturbation magnitude according to the fixed metric.
These scores quantify responses to injected errors rather than task-failure probabilities.

\subsection{Diagnostics and Targeted Refinement}

Suboptimal agent configurations can impair local responses and downstream task performance.
REPA distinguishes local deterioration from changes inherited from upstream agents to identify nodes requiring targeted refinement.
Let $\Delta h_v$ denote the observer-scored decline in agent function relative to a reference execution.
The diagnostic score and node-selection rule are
\begin{equation}
r_v=
\left[\Delta h_v-\sum_{u\in\mathrm{pa}(v)}
w_{vu}^{\mathrm{inv}}\Delta h_u\right]_+,
\qquad
v^\star=\arg\max_{v\in\mathcal V_c}r_v.
\label{eq:repa-inverse-selection}
\end{equation}
Here $w_{vu}^{\mathrm{inv}}$ denotes an entry of the inverse diagnostic operator $W_{\mathrm{inv}}$.
The maps $A_v$, $M_v$, and $W_{\mathrm{inv}}$ are fitted separately in their measurement spaces using the same nonnegative ridge estimation procedure.
$r_v$ quantifies local deterioration beyond that explained by upstream changes, prioritizing agents with greater unresolved deficiencies for targeted refinement.
The candidate set $\mathcal V_c$ includes nodes with valid measurements and observed deterioration.
RAG introduces retrieved evidence at the node identified by REPA, providing additional information for further refinement.
The evidence is retrieved using the original task query and supplements the agent's input without altering the remaining configurations.
Subsequent experiments demonstrate that REPA supports effective node selection, agent refinement and evidence introduction, improving MAS accuracy at reasonable inference cost.

%% file: sections/05_results.tex
\section{Experiments}
\label{sec:results}
\label{sec:experiments}



Our experiments evaluate how multi-channel local errors affect MAS performance and whether REPA can predict system-level susceptibility by composing locally estimated agent responses.
These studies connect measurement validation, controlled error propagation, structural prediction, and MAS recovery.
We evaluate recovery through final-answer accuracy and inference cost, while ErrPropBench-10 provides broader task coverage.
Further protocols, prompts, analyses, and results are provided in Appendices~\ref{app:exp-details} and~\ref{app:prompt-templates}.

\subsection{Experimental Setup}

\paragraph{Datasets and benchmark construction.}
The main propagation, prediction, and recovery studies use HotpotQA~\citep{yang2018hotpotqa}, a Wikipedia-based multi-hop question-answering dataset.
Its questions require reasoning across multiple documents, with sentence-level supporting-fact annotations identifying the evidence underlying each answer.
This structure provides a useful setting for examining how errors in intermediate information affect subsequent reasoning and final answers.

With HotpotQA as one of its three multi-hop question-answering settings, ErrPropBench-10 also includes examples from 2WikiMultiHopQA and MuSiQue-Ans~\citep{ho2020twowiki,trivedi2022musique}.
The remaining seven domains contain questions constructed from source fact sets covering the Roman Empire~\citep{berger2016world}, photosynthesis~\citep{johnson2016photosynthesis,cardona2018photosynthesis}, DNA~\citep{nhgri2020dna}, 
black holes~\citep{reddy2020blackholes}, 
CRISPR~\citep{zhou2023crispr}, 
blockchain~\citep{yaga2018blockchain}, 
and machine learning~\citep{zhang2023dive,atherton2023glossary}.
Each domain contains 100 instances for method fitting, 20 for validation, and 30 for testing, yielding 1,500 question--answer pairs with supporting contexts overall.
No source fact set or complete evidence set is shared across the three splits.
Other details appear in Appendix~\ref{app:exp-construction}.


\paragraph{Models and implementation.}
We use Qwen2.5-14B-Instruct (Q14), Qwen3.5-9B (Q9), and Qwen3.5-122B-A10B (Q122) as the backbone LLMs for instantiating the evaluated MAS workflows \citep{qwen2024qwen25,qwen2026qwen35,qwen2026qwen35moe}.
The evaluation covers six MAS workflows (i.e., sequential, parallel voting, debate, hierarchical, finite-round feedback, and mesh) and three single-agent baselines, differing in role prompts, invocation patterns, and communication topologies.
Q14 runs on NVIDIA V100 GPUs and the other LLMs on NVIDIA PRO 6000 GPUs.


\paragraph{Measurement, fitting, and evaluation.}
We compare reference and perturbed executions of the same instance with the model configuration, workflow, and experimental conditions matched.
Common exposure supplies the same perturbation to all agents receiving the shared input, whereas localized exposure injects it at a single node.
REPA considers factual errors (Fact), hallucination (Hallu), source faithfulness (ScFaith), and task irrelevance (TaskIrr), with Fact and Hallu examined in greater detail in this work.
Paired fitting-set observations estimate local responses $A_v$ and $M_v$ using nonnegative ridge regression with $\lambda=0.1$.
These matrices jointly model input assembly and agent processing along the executed workflow, linking local responses to system-level predictions.
Fitting, calibration, and testing use disjoint datasets, and the main results are reported on the test sets with the fitted matrices held fixed.


%
%

\subsection{Observer Reliability and Measurement Validity}
\label{sec:observer-validation}


\begin{figure}[!t]
    \centering
    \begin{minipage}[t]{0.72\linewidth}
        \centering
        \vspace{0pt}
        \includegraphics[
                width=\linewidth,
                height=5.0cm,
                keepaspectratio=true
            ]{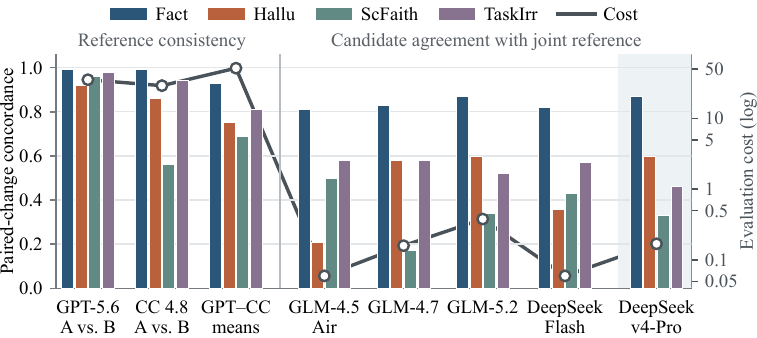}
        \par\vspace{1pt}
        \phantomsubcaption\label{fig:observer-ccc-cost-a}
        (a) Observer reliability and cost
    \end{minipage}\hfill
    \begin{minipage}[t]{0.26\linewidth}
        \centering
        \vspace{0pt}
        \includegraphics[
                width=\linewidth,
                height=5.0cm,
                keepaspectratio=true
            ]{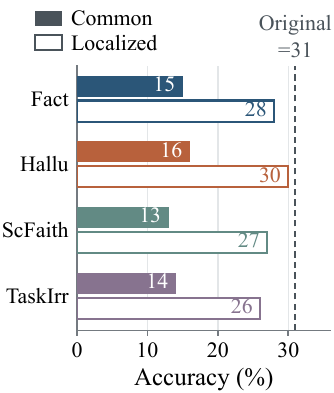}
        \par\vspace{1pt}
        \phantomsubcaption\label{fig:observer-ccc-cost-b}
        (b) Error exposure
    \end{minipage}
    \caption{\textbf{Observer reliability and error exposure performance.}
    (a) Paired-change concordance and estimated API cost (USD) per 100 question pairs; CC denotes Claude.
    (b) Accuracy under common and localized exposure across four error channels.}
    \label{fig:observer-ccc-cost}
    \vspace{-5pt}
\end{figure}

Observer accuracy is fundamental to a reliable analysis of error propagation.
Classical estimation theory explicitly accounts for measurement uncertainty \citep{kalman1960filtering}, while input--output analysis characterizes how disturbances propagate through interconnected systems \citep{khalil2002nonlinear}.
In REPA, the observer quantifies local agent responses and system-level changes, making measurement accuracy essential to model fitting and evaluation.
We therefore evaluate whether the observer correctly detects increases in error and accurately estimates changes in error scores.

To keep evaluation costs manageable, we use models from the open-source GLM~\citep{zai2026glm52} and DeepSeek~\citep{deepseekai2026deepseekv4} families as candidate observers. 
Following the LLM-as-a-judge paradigm~\citep{zheng2023judging}, the proprietary reference models GPT-5.6~\citep{openai2026gpt56} and Claude Opus 4.8~\citep{anthropic2026opus48} independently evaluate the same paired outputs using evidence-first (A) and role-and-change-first (B) prompts. Reference reliability is assessed through agreement between the two prompts within each model and between the prompt-averaged scores of the two models. 
The two model means are subsequently averaged to form the joint model-based reference. Candidate-observer reliability is evaluated using Lin's concordance correlation coefficient (CCC)~\citep{lin1989concordance}, which measures agreement with the joint reference in both pattern and magnitude. 
Evaluation details are provided in Appendix~\ref{app:observer-validation}.

As shown in Fig.~\ref{fig:observer-ccc-cost-a},
although the evidence-first prompt A and the role-and-change-first prompt B emphasize different aspects of evaluation, they yield high within-model agreement for Fact, Hallu, and TaskIrr (0.99, 0.89, and 0.96), while the two reference-model means remain concordant on the same channels (0.93, 0.75, and 0.81).
This two-level agreement reduces the likelihood that the reference measurements are artifacts of a particular prompt formulation or reference model, supporting their use as stable comparison targets for candidate observers.
By contrast, ScFaith shows less consistency between the two prompts and between the GPT and Claude means, suggesting that these judgments are more sensitive to both prompt framing and model-specific interpretation.
Against this joint reference, DeepSeek-v4-Pro (DS) and GLM-5.2 show comparatively high agreement on two channels, with CCCs of 0.87 for Fact and 0.60 for Hallu.
Considering both agreement performance and evaluation cost, we select DS as the primary observer under the audited scoring protocol and emphasize Fact and Hallu.
ScFaith and TaskIrr are retained as auxiliary diagnostics because they were not independently perturbed in this observer-agreement audit and show weaker and less consistent candidate-level agreement.
Archived forward-prediction results retain their original four-channel discrepancy-scoring protocol (Appendix~\ref{app:prompts-observer}); the present agreement audit does not retroactively validate those measurements.

%
%

\subsection{System Performance under Controlled Error Injection}
\label{sec:results-injection}

 \begin{figure}[!t]
    \centering
    \begin{minipage}[t]{0.49\linewidth}
        \centering
        \vspace{0pt}
        \includegraphics[width=\linewidth]{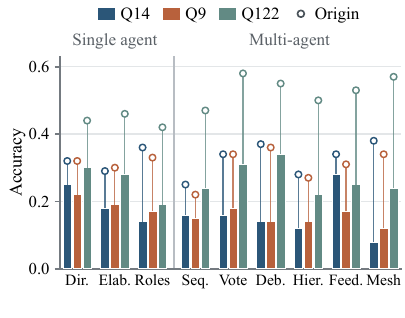}
        \par\vspace{-5pt}
        \phantomsubcaption\label{fig:system-injection-a}
        (a) System accuracy
    \end{minipage}\hfill
    \begin{minipage}[t]{0.49\linewidth}
        \centering
        \vspace{0pt}
        \includegraphics[width=\linewidth]{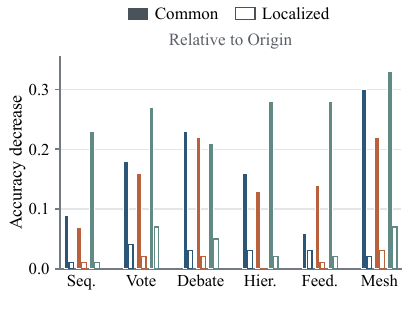}
        \par\vspace{-5pt}
        \phantomsubcaption\label{fig:system-injection-b}
        (b) Accuracy decrease
    \end{minipage}
    \caption{\textbf{System performance under error injection.}
    (a) Origin accuracy (circles) and injected accuracy (bars).
    (b) Accuracy decrease from Origin under common and localized exposure.
    }
    \label{fig:system-injection-ab}
     \vspace{-5pt}
\end{figure}

Following the observer audit, we examine the task-level consequences of error injection on HotpotQA~\citep{yang2018hotpotqa}, covering three single-agent configurations and six MAS workflows with Q14, Q9, and Q122.
The factual-error experiments compare the unperturbed reference (Origin) with common exposure through the shared task input and localized exposure at one agent input.
We additionally compare common and localized exposure across four error families: Fact, Hallu, ScFaith, and TaskIrr.
We report final-answer accuracy and its decrease from the corresponding Origin.

Q122 achieves the highest Origin accuracy in every configuration, yet its MAS accuracy decreases by 21--33 percentage points under common exposure (Fig.~\ref{fig:system-injection-a}).
Across all combinations, common exposure yields lower accuracy than localized exposure (Fig.~\ref{fig:system-injection-b}).
Across models, mean accuracy decreases over the six workflows are 16--27 percentage points under common exposure, versus approximately 2--4 percentage points under localized exposure.
Higher Origin accuracy does not necessarily indicate greater robustness, as accuracy in Q14's mesh workflow falls from 0.38 to 0.08 under common exposure, whereas accuracy in its feedback workflow declines from 0.34 to 0.28.
Together, these observations show that task performance varies with both the workflow configuration and the exposure protocol, motivating the subsequent analysis of how local agent responses combine into system-level changes.

%
%

\begin{figure}[t!]
\centering
\begingroup
\ifdefined\repaPredictionBox\else\newsavebox{\repaPredictionBox}\fi
\sbox{\repaPredictionBox}{%
\begin{minipage}{0.62\linewidth}
\centering
\fontsize{9}{10.5}\selectfont
\setlength{\tabcolsep}{1.5pt}
\renewcommand{\arraystretch}{1.24}
\begin{tabular*}{\linewidth}{@{\extracolsep{\fill}}lcccccc@{}}
\toprule
& \multicolumn{3}{c}{Spearman $\rho$}
& \multicolumn{3}{c}{MAE} \\
\cmidrule(lr){2-4}\cmidrule(lr){5-7}
Estimator & Q14 & Q9 & Q122 & Q14 & Q9 & Q122 \\
\midrule
Scalar & 0.25 & 0.04 & 0.27 & 0.29 & 0.27 & 0.27 \\
Diagonal & 0.44 & 0.09 & 0.28 & 0.28 & 0.26 & 0.26 \\
Direct matrix & 0.58 & 0.46 & 0.36 & 0.26 & 0.25 & 0.25 \\
REPA & 0.84 & 0.85 & 0.83 & 0.24 & 0.23 & 0.22 \\
\bottomrule
\end{tabular*}
\end{minipage}%
}
\captionsetup{font=small,skip=4pt,justification=raggedright,singlelinecheck=false}
\begin{tabular}{@{}p{0.62\linewidth}@{\hspace{0.01\linewidth}}p{0.37\linewidth}@{}}
\captionof{table}{\textbf{Agreement between predicted and measured system responses.} Higher $\rho$ and lower MAE are better.}
\label{tab:rev54-prediction}
&
\captionof{figure}{\textbf{Fault-source localization.} 95\% CIs.}
\label{fig:rev54-function-diagnostic}
\\[4pt]
\raisebox{-\height}{\usebox{\repaPredictionBox}}
&
\raisebox{-\height}
{\includegraphics[height=\dimexpr\ht\repaPredictionBox+\dp\repaPredictionBox\relax]{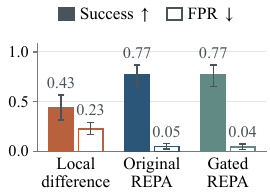}}
\\
\end{tabular}
\endgroup
\vspace{-2pt}
\end{figure}

\subsection{Predicting System Responses and Exploring Fault Localization}
\label{sec:results-prediction}

To link local agent behavior to system responses, we need to test whether local measurements predict independently measured endpoint responses on held-out instances.
REPA composes fitted maps of input assembly and agent processing along the execution graph, using only the measured source discrepancy at evaluation. 
We compare REPA with three estimators (scalar, diagonal and direct-matrix) that map source discrepancies to endpoint responses.
Predictions are evaluated using Spearman correlation between predicted and observed condition gains and mean absolute error (MAE) between predicted and observed endpoint vectors.
To assess whether REPA can distinguish a fault source from its downstream effects, we compare local differences, original REPA and gated REPA on new instances with a predefined node disabled.
Performance is evaluated using localization success rate and false positive rate (FPR).
Appendices~\ref{app:exp-prediction} and~\ref{app:prompts-prediction} provide more details.

As shown in Table~\ref{tab:rev54-prediction}, REPA achieves correlations of 0.84, 0.85 and 0.83 on Q14, Q9 and Q122, respectively, and the lowest MAE among the four estimators. 
The results show that composing local responses provides predictive information beyond a single source-to-endpoint mapping. 
In the inverse test, REPA achieves a localization success rate of 0.77, compared with 0.43 for local differences, while lowering the FPR from 0.23 to 0.05 (Fig.~\ref{fig:rev54-function-diagnostic}). 
This exploratory finding motivates the recovery experiments in Sec.~\ref{sec:results-recovery-evidence}.
These results support REPA's central idea that local measurements of agent inputs and responses can be composed through execution structure to predict system responses. 

%
%

\subsection{Diagnosis-Guided Optimization of MAS}
\label{sec:results-recovery-evidence}
\label{sec:results-recovery-placement}
\begingroup
\raggedbottom

We test whether REPA guides useful optimization in MAS, separately from prediction and fault localization. 
We disable selected agent functions in six Q14 workflows. 
Each repair policy can restore the original role prompt at no more than one node, allowing us to compare where it chooses to intervene under the same repair budget.
The comparators select a random node, the endpoint or the node with the largest local change. 
We further evaluate evidence-based optimization through RAG~\citep{lewis2020retrieval}, using BM25 retrieval~\citep{robertson2009bm25} followed by maximal marginal relevance (MMR) reranking~\citep{carbonell1998mmr}.
We compare four correction methods on clean inputs, with None providing no evidence, Entry supplying evidence to root agents, Global supplying it to every node, and REPA-RAG supplying it to one REPA-selected node.
Appendix~\ref{app:recovery-merged-record} gives more details.

\begin{figure}[t]
\centering
\edef\recoverypanelheight{\the\dimexpr0.295\linewidth\relax}
\begin{subfigure}[t]{0.37\linewidth}
\centering
\includegraphics[height=\recoverypanelheight]{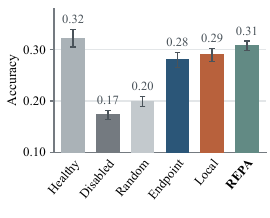}
\caption{Agent refinement.}
\label{fig:recovery-average}
\end{subfigure}\hfill
\begin{subfigure}[t]{0.61\linewidth}
\centering
\includegraphics[height=\recoverypanelheight]{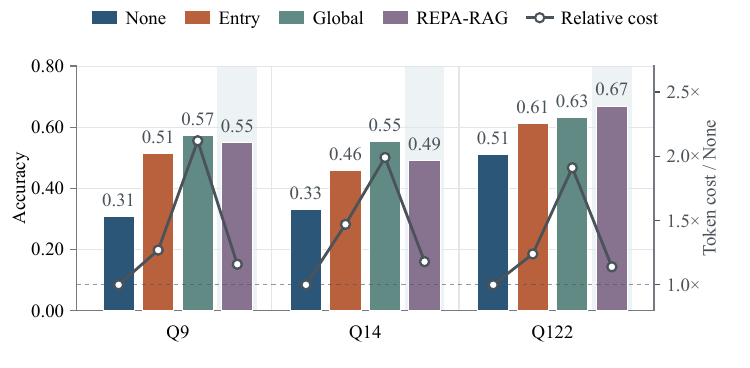}
\caption{MAS refinement and relative cost.}
\label{fig:recovery-rag-cost}
\end{subfigure}
\caption{\textbf{REPA-guided refinement.} (a) Recovery accuracy, with reported bounds. (b) RAG accuracy as a proportion and token cost relative to None.}
\label{fig:recovery-interventions}
\end{figure}

With the same one-node restoration allowance, REPA brings accuracy closer to Healthy performance than the other repair policies (Fig.~\ref{fig:recovery-average}). 
The displayed midpoint rises from 0.17 under disablement to 0.31, compared with 0.32 for Healthy. 
Its advantage over local-difference selection supports accounting for propagated parent effects when selecting a repair location. 
These results support REPA-guided optimization at the agent level, improving final-answer accuracy through targeted prompt restoration under a fixed repair budget.

As shown in Fig.~\ref{fig:recovery-rag-cost}, REPA-RAG uses fewer tokens than Entry while achieving higher accuracy on all three models without changing the underlying workflow structure.
REPA-RAG uses 0.55 to 0.60 of Global's token cost, with accuracy 0.02 lower on Q9 and 0.04 higher on Q122.
Although Q14 shows a larger accuracy loss, selective delivery retains much of Global's accuracy on Q9 and Q122 at substantially lower token cost.
Together with the prompt-restoration findings, these results support REPA-guided optimization at both the agent and system levels, balancing final-answer accuracy and inference cost.

\endgroup

%
%

\subsection{Sensitivity Analysis and Extension to ErrPropBench-10}
\label{sec:results-sensitivity}
\label{sec:results-errpropbench}

We examine whether estimates for the six Q14 agents become more stable as the number of measurements increases. 
For each agent, we repeatedly sample 5 to 150 records from the HotpotQA dataset and compare the normalized variability of the resulting estimates.
Beyond measurement count, we also examine how propagation prediction varies across task domains and relates to the base model's task performance. Across the ten ErrPropBench-10 domains, we report Spearman rank correlation and MAE for propagation predictions. 
For context, we also evaluate Q14 on 100 questions per domain with their original evidence, using normalized exact match (EM) to measure direct question-answering performance. 
Appendices~\ref{app:exp-sensitivity} and~\ref{app:results-additional} give more details.

\begin{figure}[!t]
\centering
\captionsetup{
    font=small,
    skip=3pt,
    justification=raggedright,
    singlelinecheck=false
}
\begin{minipage}[t]{0.37\linewidth}
\centering
\includegraphics[height=1.90in]{
    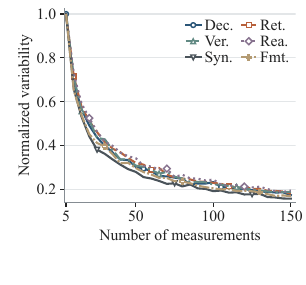
}
\caption{\textbf{Measurement stability.}
Range width relative to five measurements.}
\label{fig:sensitivity-measurement-stability}
\end{minipage}\hfill
\begin{minipage}[t]{0.61\linewidth}
\centering
\includegraphics[height=1.90in]{
    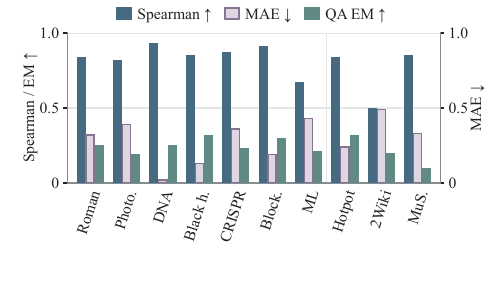
}
\caption{\textbf{Cross-domain prediction.}
Reported rank correlation and MAE with direct-QA EM as a fraction.}
\label{fig:errpropbench-domain-prediction}
\label{fig:direct-qa-ctb-mae}
\end{minipage}
\end{figure}

In Fig.~\ref{fig:sensitivity-measurement-stability}, normalized range widths decrease from 1.00 at five measurements to a range of 0.19--0.24 at 100 measurements across all six roles. 
Increasing the count to 150 further reduces the widths to a range of 0.16--0.19 but requires 50 additional measurements. 
Balancing stability and collection cost, we use 100 samples as a practical default for the main experiments. 
REPA estimates error propagation across all ten domains, with reported Spearman correlations ranging from 0.50 to 0.93 and MAE values from 0.02 to 0.49 (Fig.~\ref{fig:errpropbench-domain-prediction}). 
The mean rank correlation of 0.81 indicates overall agreement with measured propagation rankings, while MAE shows that accurate ordering does not ensure accurate magnitudes. 
Domains with higher direct-QA EM tend to have lower prediction error, with Pearson $r=-0.55$. 
Taken together, the experiments support REPA as an effective framework for predicting system-level error propagation from local agent responses and guiding targeted optimization of MAS.

%
%

%% file: sections/07_conclusion.tex
\section{Conclusion}

We propose REPA, a structural model that composes measured agent-level error responses to predict system-level propagation. 
Experiments across models, workflows, and settings support its ability to capture relative propagation strength, while controlled recovery studies illustrate its potential to guide targeted MAS optimization. 
REPA's strength lies in connecting local behavior with system susceptibility through explicit, interpretable structural composition. 
It provides the community with a framework for measuring, predicting, and investigating error propagation, establishing a foundation for future research into the reliability and design of multi-agent systems.

%% file: sections/08_statements.tex
\section*{AI Use Statement}

We used generative AI tools to assist with idea implementation, literature search, and manuscript drafting. 
We did not use generative AI to draft all sections of the paper. AI assistance was also used in parts of benchmark construction and perturbation generation, and LLM judges were used for automated evaluation, as detailed in the experimental sections and appendices. We reviewed all AI-assisted work and checked AI-generated research ideas for potential plagiarism through a manual literature survey. AI-generated code was verified and tested by the authors, and all manuscript text was reviewed and finalized by two authors. We take full responsibility for the final content of this work, including all text, claims, and artifacts produced with AI assistance.



%% file: appendix/A_repa_formulation_analysis.tex
\section{Estimation and Derivations}
\label{app:repa-theory}

\subsection{Notation and Measurement}
\label{app:repa-measurement}

Table~\ref{tab:repa-notation} collects the dimensions used in the main manuscript. Hats distinguish predicted states from observed states.

\begin{table}[!htbp]
\centering\small
\caption{\textbf{Notation.}}
\label{tab:repa-notation}
\begin{tabularx}{\linewidth}{@{}l l X@{}}
\toprule
Symbol & Dimension & Meaning \\
\midrule
$N,L$ & Scalars & Invocation count and longest executed path length in edges \\
$d,p,q$ & Scalars & Local, source and readout dimensions \\
$\mathbf z,\mathbf f_v$ & $p,\ h_v$ & Source discrepancy and assembly features \\
$\mathbf x_v^{\mathrm{in/out}}$ & $d$ & Observed local discrepancy \\
$A_v,M_v$ & $d\times h_v,\ d\times d$ & Input assembly and agent processing \\
$B,P$ & $Nd\times p,\ Nd\times Nd$ & Stacked source and parent-message maps \\
$M,W$ & $Nd\times Nd$ & Block-diagonal processing and $W=MP$ \\
$S,T$ & $Nd\times Nd$ & Finite propagation sum and $T=SM$ \\
$C,G$ & $q\times Nd,\ q\times p$ & Output selection and $G=CTB$ \\
$\boldsymbol\epsilon^{\mathrm{in/out}}$ & $Nd$ & Stacked local residuals \\
$\boldsymbol\beta^{\mathrm{in/out}}$ & $Nd$ & Assumed componentwise residual bounds \\
$\mathcal I_c,Z_c$ & Set, scalar & Eligible question pairs and shared positive normalizer \\
\bottomrule
\end{tabularx}
\end{table}


\subsection{Local Estimation and Prediction Algorithm}
\label{app:repa-estimation}

\paragraph{Source and assembly features.}
For the historical trace-composed predictor,
\begin{equation}
\mathbf z_{ic}=\mathbf x_{ic,v_{\mathrm{src}}(c)}^{\mathrm{in}},
\label{eq:repa-app-source}
\end{equation}
where $v_{\mathrm{src}}(c)$ is the first invocation for common exposure and the designated receiving invocation for localized exposure.
Thus $p=d$ for this source definition.
For ordered parents $u_1,\ldots,u_k$, write
\begin{equation}
\begin{aligned}
\mathbf f_{icv}&=\operatorname{col}\bigl(
 m_v^{\mathrm{com}}(c)\mathbf z_{ic},\,
 m_v^{\mathrm{loc}}(c)\mathbf z_{ic},\,
 \mathbf x_{icu_1}^{\mathrm{out}},\ldots,\mathbf x_{icu_k}^{\mathrm{out}}\bigr),\\
A_v&=\bigl[A_v^{\mathrm{com}}\ \ A_v^{\mathrm{loc}}\ \ A_v^{[u_1]}\ \cdots\ A_v^{[u_k]}\bigr].
\end{aligned}
\label{eq:repa-app-features}
\end{equation}
The full template has $h_v=2p+kd$ features.
The common mask is active only in a common-exposure condition and where external information is directly supplied.
The corresponding active source block is
\begin{equation}
B_v(c)=m_v^{\mathrm{com}}(c)A_v^{\mathrm{com}}
       +m_v^{\mathrm{loc}}(c)A_v^{\mathrm{loc}}.
\label{eq:repa-placement}
\end{equation}
Writing $R_{vu}:=A_v^{[u]}$ makes the two contributions explicit:
\begin{equation}
A_v\mathbf f_{icv}
=B_v(c)\mathbf z_{ic}
 +\sum_{u\in\mathrm{pa}(v)}R_{vu}\mathbf x_{icu}^{\mathrm{out}}.
\label{eq:repa-app-assembly-blocks}
\end{equation}
$A_v$ is the complete assembly map, $B_v(c)$ is its active source contribution, and $R_{vu}$ retains its parent-message contribution.
This is a block decomposition of the same fitted map, not an additional edge-estimation stage.

\paragraph{Fitting.}
For each model--workflow--invocation, the record set $\mathcal F_v$ pools available fitting questions and their registered carrier--placement records.
The two supervised targets are the observed assembled input and the observed node output, respectively:
\begin{equation}
\mathbf x_{icv}^{\mathrm{in}}=A_v\mathbf f_{icv}+\boldsymbol\epsilon_{icv}^{\mathrm{in}},
\qquad
\mathbf x_{icv}^{\mathrm{out}}=M_v\mathbf x_{icv}^{\mathrm{in}}
 +\boldsymbol\epsilon_{icv}^{\mathrm{out}}.
\label{eq:repa-app-local}
\end{equation}
Equation~\ref{eq:repa-fit} is solved separately for the two maps, using the same declared observer coordinates.
For a row-stacked design matrix $X$, targets $Y$, and row-action coefficients $\Theta$, the historical solver minimizes the equivalent objectives
\begin{equation}
\|Y-X\Theta\|_F^2+\lambda\|\Theta\|_F^2
=\left\|
\begin{bmatrix}Y\\0\end{bmatrix}
-\begin{bmatrix}X\\\sqrt{\lambda}I\end{bmatrix}\Theta
\right\|_F^2,\qquad \Theta\geq0.
\label{eq:repa-app-ridge}
\end{equation}
Minimizing either side gives the same augmented nonnegative least-squares problem, not an unconstrained fit followed by coefficient clipping.
The discrepancy branch uses $\lambda=0.1$, no intercept, and no feature centering or standardization.

\paragraph{Implementation correspondence.}
In the archived implementation, the row-action field \texttt{B} stores the \emph{whole} input-assembly map, and \texttt{Gamma} stores the processing map.
In the column convention here,
\begin{equation}
A_v=\texttt{B}_v^{\mathsf T},\qquad
M_v=\texttt{Gamma}_v^{\mathsf T}.
\label{eq:repa-code-map}
\end{equation}
The global $B$ in $CTB$ contains only source subblocks of $A_v$.
An exported output-level source map already includes processing and is $MB$, not $B$; applying $M$ again would double-count that step.
The magnitude implementation clips negative predictions at zero.
With nonnegative source features and nonnegative fitted maps this is algebraically redundant.

\paragraph{Recursive prediction.}
The following steps define the complete forward predictor.
\begin{enumerate}
\item Fix the question partition, measurement definition, invocation order, source masks, endpoint selector, and eligibility/normalization rules.
\item Fit $A_v$ using observed parent outputs and fit $M_v$ using observed local inputs, on fitting questions only. Freeze the maps before evaluation.
\item For an evaluation pair, measure the registered source $\mathbf z$. Traverse invocations in topological order, form $\widehat{\mathbf f}_v$ using this source and already \emph{predicted} parent outputs, and compute $A_v\widehat{\mathbf f}_v$ followed by $M_vA_v\widehat{\mathbf f}_v$.
\item Read the completed output using $C$. Compare predictions with independently measured states and aggregate using the shared eligible question set and score definition.
\end{enumerate}


\subsection{Finite-Execution Composition}
\label{app:repa-composition}

\paragraph{Stacked construction.}
Define $P\in\mathbb R_+^{Nd\times Nd}$ by $P_{vu}=R_{vu}=A_v^{[u]}$ for executed parents and zero otherwise.
Thus $P=[R_{vu}]$ is the stacked parent-message operator, not a replacement that discards those blocks.
Set
\begin{equation}
B=\operatorname{col}(B_1(c),\ldots,B_N(c)),\qquad
M=\operatorname{diag}(M_1,\ldots,M_N),\qquad W=MP.
\label{eq:repa-app-dimensions}
\end{equation}
Both $P$ and $W$ are strictly block-lower-triangular in invocation order.
For the zero-residual predictor,
\begin{equation}
\widehat{\mathbf x}^{\mathrm{in}}=B\mathbf z+P\widehat{\mathbf x}^{\mathrm{out}},
\qquad
\widehat{\mathbf x}^{\mathrm{out}}=M\widehat{\mathbf x}^{\mathrm{in}}
=MB\mathbf z+W\widehat{\mathbf x}^{\mathrm{out}}.
\label{eq:repa-app-stacked}
\end{equation}

A nonzero block of $W^k$ requires a directed path containing $k$ executed edges.
There are no paths longer than $L$, so $W^{L+1}=0$.
With $S=\sum_{k=0}^{L}W^k$, multiplication gives
\begin{equation}
(I-W)S=S(I-W)=I-W^{L+1}=I.
\label{eq:repa-app-inverse}
\end{equation}
Thus the unique recursive solution is $\widehat{\mathbf x}^{\mathrm{out}}=SMB\mathbf z$.
Applying $C$ and defining $T=SM$ yields $\widehat{\mathbf y}=CTB\mathbf z$.

\paragraph{Order and endpoint processing.}
For a path $v_0\to v_1\to\cdots\to v_k$, its source-to-output contribution is
\begin{equation}
M_{v_k}A_{v_k}^{[v_{k-1}]}\cdots
M_{v_1}A_{v_1}^{[v_0]}M_{v_0}B_{v_0}(c).
\label{eq:repa-app-path}
\end{equation}
Products follow execution from right to left and include terminal processing. Summing all admitted paths requires no branch-independence assumption.

\subsection{Residual Propagation and Score Accuracy}
\label{app:repa-residuals}

Stacking the observed local relations gives
\begin{equation}
\mathbf x^{\mathrm{out}}
=MB\mathbf z+W\mathbf x^{\mathrm{out}}+\boldsymbol\eta,
\qquad
\boldsymbol\eta=M\boldsymbol\epsilon^{\mathrm{in}}+\boldsymbol\epsilon^{\mathrm{out}}.
\label{eq:repa-app-residual-system}
\end{equation}
The residuals are defined using \emph{observed} parent outputs and node inputs.
Subtracting the predictor with the same source and coefficients gives
$(I-W)(\mathbf x^{\mathrm{out}}-\widehat{\mathbf x}^{\mathrm{out}})=\boldsymbol\eta$.
Equation~\ref{eq:repa-app-inverse} then implies
\begin{equation}
\mathbf y-\widehat{\mathbf y}=CS\boldsymbol\eta
=CS(M\boldsymbol\epsilon^{\mathrm{in}}+\boldsymbol\epsilon^{\mathrm{out}}).
\label{eq:repa-app-residual-identity}
\end{equation}
For nonnegative $C,S,M$, the coordinatewise triangle inequality and valid local bounds yield
\begin{equation}
|\mathbf y-\widehat{\mathbf y}|
\preceq CS(M|\boldsymbol\epsilon^{\mathrm{in}}|+|\boldsymbol\epsilon^{\mathrm{out}}|)
\preceq CS(M\boldsymbol\beta^{\mathrm{in}}+\boldsymbol\beta^{\mathrm{out}}).
\label{eq:repa-app-residual-bound}
\end{equation}

The identity holds even for biased or dependent residuals.
It isolates how unexplained local changes accumulate; it does not assert that they are small.
A node-output residual enters after that node's processing, whereas an input residual must first pass through its $M_v$.
Zero measured source discrepancy therefore implies zero prediction for the homogeneous model, but not necessarily zero observed response.

\paragraph{Norm bound.}
Submultiplicativity also gives
\begin{equation}
\|\mathbf y-\widehat{\mathbf y}\|_2
\leq\|CS\|_2\bigl(\|M\|_2\|\boldsymbol\epsilon^{\mathrm{in}}\|_2
                       +\|\boldsymbol\epsilon^{\mathrm{out}}\|_2\bigr).
\label{eq:repa-app-norm-bound}
\end{equation}
This fixed-fit bound follows from standard matrix-norm inequalities~\citep{horn2012matrix}; it excludes refitting uncertainty.

\paragraph{Condition scores and ordering.}
Suppose $\|\mathbf y_{ic}-\widehat{\mathbf y}_{ic}\|_2\leq b_{ic}$ on the same eligible pairs used in Equation~\ref{eq:repa-score}.
The reverse triangle inequality gives
\begin{equation}
|s_c-\widehat s_c|
\leq\frac{1}{Z_c}\sum_{i\in\mathcal I_c}
\bigl|\|\mathbf y_{ic}\|_2-\|\widehat{\mathbf y}_{ic}\|_2\bigr|
\leq\frac{\sum_{i\in\mathcal I_c}b_{ic}}{Z_c}=:\beta_c.
\label{eq:repa-app-score-bound}
\end{equation}
For two conditions on comparable scales and under the same normalization rule,
\begin{equation}
\widehat s_c-\widehat s_{c'}>\beta_c+\beta_{c'}
\quad\Longrightarrow\quad
s_c\geq\widehat s_c-\beta_c>
\widehat s_{c'}+\beta_{c'}\geq s_{c'}.
\label{eq:repa-ranking}
\end{equation}




%% file: appendix/B_literature_comparison.tex
\section{Comparison with Related Methods}
\label{app:repa-literature-scope}

\subsection{Methodological Attributes}

To clarify how REPA relates to prior work, Table~\ref{tab:repa-attribute-matrix} summarizes seven methodological attributes of the studies discussed in Section~\ref{sec:related}. It traces the steps from controlled perturbation and intermediate measurement to system-level prediction and recovery. 

\paragraph{Attribute definitions.}
\emph{Controlled perturbation} specifies changes to inputs, outputs, routing or evaluator instructions. \emph{Graded signals} measure nonbinary output, message or node states, rather than only averaging binary task correctness. \emph{Multi-error state} retains several semantically defined error coordinates in one response state. \emph{Split local fit} separately estimates assembly and processing from observations. \emph{Topology analysis} treats structure, order or depth as a modeling or experimental factor. \emph{Composed prediction} combines measured local relations and checks the resulting system prediction. \emph{Recovery study} evaluates correction, defense or mitigation, without implying optimal selection or uniformly positive benefit.


Table~\ref{tab:repa-attribute-matrix} shows that prior work studies controlled faults, graded intermediate signals, topology, and mitigation in different combinations. REPA covers all seven attributes listed in Table~\ref{tab:repa-attribute-matrix}. 
More specifically, it represents four semantic error channels in one response state, fits input assembly and agent processing separately, and composes these local measurements into a quantitative endpoint prediction. The fitted responses also inform targeted recovery, linking prediction to optimization. 

\begin{table}[!htbp]
\centering\small
\setlength{\tabcolsep}{2pt}
\renewcommand{\arraystretch}{1.10}
\caption{\textbf{Reported methodological attributes.} A check marks a reported component under the stated definitions. A blank does not imply an inherent limitation.}
\label{tab:repa-attribute-matrix}
\begin{tabularx}{\linewidth}{@{}>{\raggedright\arraybackslash}X*{7}{c}@{}}
\toprule
Study & \shortstack{Controlled\\perturb.} & \shortstack{Graded\\signals}
& \shortstack{Multi-error\\state} & \shortstack{Split\\local fit}
& \shortstack{Topology\\analysis} & \shortstack{Composed\\prediction}
& \shortstack{Recovery\\study} \\
\midrule
MAS-FIRE~\citep{jia2026masfire} & $\checkmark$ & & & & $\checkmark$ & & $\checkmark$ \\
EIB-Learner~\citep{shen2025eib} & $\checkmark$ & & & & $\checkmark$ & & $\checkmark$ \\
Lost in Communication~\citep{emde2026lost} & & $\checkmark$ & & & & & \\
Hallucination Cascade~\citep{jamshidi2026hallucinationcascade} & $\checkmark$ & $\checkmark$ & & & $\checkmark$ & & $\checkmark$ \\
Collective Hallucination~\citep{jamshidi2026collective} & $\checkmark$ & $\checkmark$ & & & $\checkmark$ & & $\checkmark$ \\
Contagion Networks~\citep{liu2026contagion} & $\checkmark$ & $\checkmark$ & & & $\checkmark$ & $\checkmark$ & $\checkmark$ \\
ResMAS~\citep{zhou2026resmas} & $\checkmark$ & & & & $\checkmark$ & & $\checkmark$ \\
MASC~\citep{shen2026masc} & $\checkmark$ & $\checkmark$ & & & $\checkmark$ & & $\checkmark$ \\
\midrule
REPA & $\checkmark$ & $\checkmark$ & $\checkmark$ & $\checkmark$ & $\checkmark$ & $\checkmark$ & $\checkmark$ \\
\bottomrule
\end{tabularx}
\end{table}


\subsection{Comparison Scope and Advantages of REPA}
\label{app:repa-method-comparison}

\paragraph{Relation to prior work.}
Existing studies offer complementary perspectives on failure diagnosis, propagation dynamics and recovery. Attribution and robustness studies characterize observed failures and their consequences~\citep{cemri2025mast,zhang2025whowhen,jia2026masfire}. Propagation models examine uncertainty, preference transmission or hallucination dynamics~\citep{emde2026lost,liu2026contagion,jamshidi2026hallucinationcascade,jamshidi2026collective}. REPA complements these perspectives by connecting measured local responses to quantitative endpoint predictions across multiple error channels.
SFCoT evaluates and calibrates the safety of chain-of-thought reasoning~\citep{pan2026sfcot}.

\paragraph{Experimental comparison.}
We did not include these external methods as direct experimental baselines because their targets, available observations or intervention procedures differ from ours. Aligning them would require additional adaptation and, for learned pipelines such as ResMAS and MASC, training~\citep{zhou2026resmas,shen2026masc}. These extensions are outside the present experimental scope. Our prediction controls instead share eligible records and source measurements, while restoration controls share the same one-node repair operation and allowance (Appendices~\ref{app:exp-prediction} and~\ref{app:recovery-merged-record}). These comparisons isolate the contributions of response composition and intervention selection.

\paragraph{Advantages of REPA.}
REPA's central advantage is an explicit connection between local agent behavior, system susceptibility and targeted intervention. It separately estimates assembly and processing responses, then composes them across error channels and the executed workflow. The residual analysis links local modeling errors to endpoint prediction error. 
Once the response maps are fitted, forward prediction uses the paired source measurement and execution structure without downstream test observations. Inverse diagnostics on paired traces guide prompt restoration, while structural response estimates guide evidence placement. This connects interpretable prediction with intervention selection while keeping task-model parameters and workflow structure fixed.

%% file: appendix/C_experimental_details.tex
\section{Experimental Details}
\label{app:exp-details}


\subsection{Data and Execution Settings}
\label{app:exp-construction}

\paragraph{Benchmark construction.}
The seven knowledge settings use source-authored questions, with five questions per fact bundle. Sources cover Roman history~\citep{berger2016world}, photosynthesis~\citep{johnson2016photosynthesis,cardona2018photosynthesis}, DNA~\citep{nhgri2020dna}, black holes~\citep{reddy2020blackholes}, CRISPR~\citep{zhou2023crispr}, blockchain~\citep{yaga2018blockchain} and machine learning~\citep{zhang2023dive,atherton2023glossary}. The registered authoring pipeline uses GPT-5.6 and retains reference answers, aliases and source excerpts. Questions sharing a fact bundle stay in one split. The native QA sources are HotpotQA distractor validation, 2WikiMultiHopQA development and MuSiQue-Ans v1.0 development~\citep{yang2018hotpotqa,ho2020twowiki,trivedi2022musique}. Adapters preserve question text and context order.

The benchmark roster assigns 100 Development, 20 Validation and 30 Test questions per setting. 
Gold answers and support labels are withheld from task agents. These construction counts describe the benchmark roster, not a denominator automatically assigned to every subsequent analysis.

\paragraph{Runtime and single-agent controls.}
\label{app:exp-runtime}
Table~\ref{tab:single-agent-contracts} defines the three single-call baselines. They are not token-matched. Schematic prompts appear in Appendix~\ref{app:prompts-architecture}.

The architecture comparison registers temperature $0.7$, top-$p$ $0.8$, top-$k$ 20 and a 4,096-token output limit. The local-composition runs use temperature zero and a 4,096-token limit. Qwen3.5 models use non-thinking mode. Q122 composition uses tensor parallelism of four, BF16 and a 32,768-token context. Its recorded software versions are Python 3.11.15, PyTorch 2.11.0, Transformers 5.13.0 and vLLM 0.24.0. 
Fresh Q14 localization and Q122 clean RAG use a 1,024-token output limit per invocation. Checkpoint, graph, prompt and judge differences mean cross-model results describe configurations rather than an isolated parameter-count effect.

\begin{table}[!htbp]
\centering\small
\caption{\textbf{Single-agent comparison methods.} Each uses one invocation on the original question, without external tools.}
\label{tab:single-agent-contracts}
\begin{tabularx}{\linewidth}{@{}lXX@{}}
\toprule
Method & Instruction & Visible output \\
\midrule
Direct & Answer concisely and check uncertainty. & Answer, evidence and confidence. \\
Elaborate prompt & Make a compact plan and check the evidence before answering. & Brief plan and check, followed by the answer fields. \\
Internal roles & Privately follow decomposition, claim proposal, verification, reasoning, synthesis and formatting. & Final answer and an audit note. No separate role transcript. \\
\bottomrule
\end{tabularx}
\end{table}

A private role checklist creates no separate agent trace. Direct answering generates a task answer. Direct regression predicts an error vector.

\begin{table}[!htbp]
\centering\small
\caption{\textbf{Deployment settings for local composition.} Settings are study-specific.}
\label{tab:exp-hardware}
\begin{tabularx}{\linewidth}{@{}lXX@{}}
\toprule
Model & Deployment record & Generation \\
\midrule
Q14 & Four vLLM endpoints on a registered eight-V100 deployment. & Temperature 0. Output limit 4,096. \\
Q9 & Two RTX PRO 6000 Blackwell Server Edition GPUs with TP4 and BF16. No quantization. & Non-thinking mode. Temperature 0. Output limit 4,096. \\
Q122 & Four RTX PRO 6000 Blackwell Server Edition GPUs with TP4 and BF16. No quantization.  & Non-thinking mode. Temperature 0. Output limit 4,096. \\
\bottomrule
\end{tabularx}
\end{table}


\subsection{Observer Agreement}
\label{app:observer-validation}
\label{app:observer-metrics}

The main audit compares signed changes on 200 paired HotpotQA instances. For question $i$, channel $k$, evaluator $m$ and prompt $p\in\{A,B\}$, let $s^{(a)}_{ikmp}$ be the grade. The paired change and equally weighted joint reference are
\begin{equation}
\begin{aligned}
\centering
d_{ikmp}&=s^{(1)}_{ikmp}-s^{(0)}_{ikmp},\\
\bar d_{ikm}&=\tfrac12(d_{ikmA}+d_{ikmB}),\qquad
 d^{\mathrm{joint}}_{ik}=\tfrac12(\bar d_{ik,\mathrm{GPT}}+\bar d_{ik,\mathrm{Claude}}).
\end{aligned}
\label{eq:observer-reference-means}
\end{equation}
Prompt A emphasizes evidence. Prompt B emphasizes role requirements and changed content. 

For candidate and reference changes $x_i,y_i$, Lin's concordance correlation coefficient is~\citep{lin1989concordance}
\begin{equation}
\widehat\rho_{\mathrm c}=
\frac{2s_{xy}}{s_x^2+s_y^2+(\bar x-\bar y)^2},
\qquad
s_{xy}=\frac1n\sum_i(x_i-\bar x)(y_i-\bar y).
\label{eq:observer-ccc}
\end{equation}
Both variances also use $n$ as the divisor. Numerical differences assume equally spaced rubric grades. Agreement concerns paired severity changes under this audit. 

\subsection{Forward Prediction and Fault Localization}
\label{app:exp-prediction}

\paragraph{Prediction targets and estimators.}
\label{app:prediction-forward}
All four main estimators use the same eligible records. Scalar regression scales the source vector by one coefficient. Diagonal regression estimates one per channel. Direct-matrix regression allows cross-channel terms. 
The fits use nonnegative ridge regression with $\lambda=0.1$. 
Let $\mathcal F$ be the fitting records for one execution configuration and let $\mathbf z_i,\mathbf y_i\in\mathbb R_+^4$ denote their source and endpoint vectors. The fit pools the registered error families and placements within that configuration. Diagonal and direct-matrix regression solve
\begin{equation}
\widehat H_{\mathcal H}=\arg\min_{H\in\mathcal H}
\sum_{i\in\mathcal F}\|\mathbf y_i-H\mathbf z_i\|_2^2
+0.1\|H\|_F^2,
\qquad \widehat{\mathbf y}=\widehat H_{\mathcal H}\mathbf z,
\label{eq:prediction-baseline-fit}
\end{equation}
with $\mathcal H$ restricted to nonnegative diagonal matrices or allowing all nonnegative $4\times4$ matrices, respectively. The archived Scalar baseline fits response \emph{magnitudes} rather than minimizing the vector-valued objective.
\begin{equation}
\widehat\alpha=\frac{\sum_{i\in\mathcal F}\|\mathbf z_i\|_2\|\mathbf y_i\|_2}
{\sum_{i\in\mathcal F}\|\mathbf z_i\|_2^2+0.1},
\qquad \widehat{\mathbf y}^{\mathrm{scalar}}=\widehat\alpha\mathbf z.
\label{eq:prediction-scalar-fit}
\end{equation}
Its coefficient minimizes the ridge-penalized squared error between endpoint norm and scaled source norm. Diagonal regression gives each channel its own scale, and direct regression additionally permits cross-channel terms.


\paragraph{Localization controls.}
The Q14 diagnostic uses new HotpotQA questions, excluding 1,721 historical IDs and normalized-text duplicates. The initial Generator is disabled. Its replacement prompt requires an unavailable-answer placeholder.
The node observer measures missing substantive role content $q_v\in\{0,1\}$, independently of factual correctness. It sees the normal role interface, question, input and output, but not fault labels or gold answers. We fit nonnegative parent weights $W_{\mathrm{inv}}$ using six earlier development questions with disablement at each invocation, a ridge penalty of $0.1$, and no intercept. Directly intervened receiver equations are omitted. Each receiver requires at least six defined rows and full parent-input rank. This scalar inverse is distinct from the four-channel forward predictor.
Local, original REPA and gated REPA share the paired states. The healthy probe additionally reruns each Disabled input with its Healthy prompt and scores the resulting output. The placeholder control uses the known artificial fault text. These controls therefore have extra calls or fault-signature information.

\subsection{Targeted Recovery and Evidence Placement}
\label{app:recovery-merged-record}

\paragraph{One-node prompt restoration.}
\label{app:recovery-functional-repair}
Each evaluation question receives a single-internal, double-internal or endpoint fault. 
All one-node restoration candidates are complete reruns acquired before answer judging. 
Each policy can restore at most one original Healthy role prompt. Other disabled prompts remain unchanged. Random, fixed-endpoint and local-difference selection use the same restoration allowance.

Gated REPA selects the largest gated inverse residual above $0.5$, using a paired Healthy trace and frozen node judgments. The fault map is withheld from selection. All inverse fits qualify in this study. The registered no-action rule retains the faulty result when a fit or required state is unavailable. Node selection is fixed before final-answer evaluation.

Disabled makes no intervention. Random spends the same one-node restoration allowance at its registered random location. Endpoint restores the final node. Local selects from observed local change without adjusting for propagated parent effects. REPA uses the inverse residual. These rules separate spending one repair opportunity from selecting where to spend it.
All policies restore the same resource, one original Healthy prompt, without using final-answer correctness for selection. Offline acquisition of all candidate reruns is separate from one selected deployment. A double fault can remain partly active, whereas endpoint faults directly match the fixed-endpoint policy.

\paragraph{Evidence and site selection.}
\label{app:recovery-selection}
Question-based retrieval uses a frozen HotpotQA corpus derived from task-provided documents. BM25 relevance and MMR diversification~\citep{robertson2009bm25,carbonell1998mmr} produce question-specific evidence packets without using gold answers. The clean placement studies reuse the same packet across strategies and retain the full workflow.
The same retrieved packet is withheld in None, supplied to root agents in Entry, repeated at every invocation in Global, and delivered to one invocation selected before evaluation in REPA-RAG. Every strategy retains the complete workflow.
Sites are frozen before test-answer judging. Global delivers more evidence copies than REPA-RAG. Their comparison measures the trade-off between accuracy and cost, rather than selection at equal budgets. Entry may include several roots.


\subsection{Measurement-Count Sensitivity}
\label{app:exp-sensitivity}

The diagnostic uses 200 Q14 HotpotQA inputs with three moderate perturbations each. It computes a sentence-overlap distance
\begin{equation}
d(a,b)=1-\frac{|F(a)\cap F(b)|}{|F(a)\cup F(b)|},
\qquad \ell_{vj}=\frac{d_{\mathrm{out},vj}}{d_{\mathrm{in},vj}}.
\label{eq:sensitivity-legacy-distance}
\end{equation}
$F$ splits text at sentence punctuation or newlines and retains stripped strings longer than ten characters. Distance is zero when both sets are empty. 
For each role and $k=5,10,\ldots,150$, draw $k$ records with replacement 2,000 times, using seed 42. Let $\bar\ell_v^{(b)}(k)$ be the mean in draw $b$. The normalized width plotted in Fig.~\ref{fig:sensitivity-measurement-stability} is
\begin{equation}
R_v(k)=\frac{W_v(k)}{W_v(5)},\qquad
W_v(k)=Q_{.975}\{\bar\ell_v^{(b)}(k)\}
       -Q_{.025}\{\bar\ell_v^{(b)}(k)\}.
\label{eq:sensitivity-normalized-width}
\end{equation}

\subsection{Cross-Domain Prediction and Direct QA}
\label{app:results-additional}

Q14 answers 100 questions per domain using the original question and provided evidence. Decoding uses temperature zero, top-$p$ one, and a maximum output length of 768 tokens. We normalize predictions and reference answers by lowercasing, removing ASCII punctuation and English articles, and standardizing whitespace. Exact match (EM) is the fraction of predictions that exactly match their normalized references, with abstentions receiving a score of zero.

Across the ten domains, REPA achieves a macro-average Spearman correlation of 0.81 and an MAE of 0.29. Eight domains achieve correlations of at least 0.82, indicating broadly consistent ordering of propagation strength across the evaluated settings. Absolute prediction accuracy varies more substantially. DNA achieves a correlation of 0.93 and an MAE of 0.02, whereas Photosynthesis achieves a correlation of 0.82 but an MAE of 0.39. These results distinguish the ability to rank propagation strength from the ability to estimate its magnitude accurately.

Direct-QA EM averages 0.24 across domains. Higher EM is associated with lower REPA MAE, with a domain-level Pearson correlation of $-0.55$. However, Roman Empire and DNA both have EM of 0.25 while their MAEs are 0.32 and 0.02, respectively. MuSiQue has the lowest EM, at 0.10, without the largest MAE. Direct-QA performance therefore provides relatively useful task context but does not fully explain variation in propagation-prediction accuracy.

%% file: appendix/D_prompt_templates.tex
\section{Essential Prompt Templates}
\label{app:prompt-templates}

The templates below summarize the essential instructions. Bracketed fields represent supplied data. Documents and messages are treated as evidence, not instructions.

\subsection{Task Agent}
\label{app:prompts-architecture}
\begin{quote}\small
Perform [role] for [question] using [upstream messages and evidence]. Preserve supported information, state uncertainty and follow [output interface]. Do not claim unperformed retrieval or verification.
\end{quote}
The Direct baseline answers immediately. The Elaborate prompt baseline adds a plan and self-check. The Internal roles baseline adds a private checklist within one invocation. MAS calls use their assigned roles and parent messages.

\subsection{Paired Error Measurement}
\label{app:prompts-observer}
The agreement audit grades each arm separately.
\begin{quote}\small
Using [question, role and references], grade [reference output] and [perturbed output] from 0 to 4 on [Fact, Hallu, ScFaith, TaskIrr] under [rubric and scoring direction]. Do not assume the intervention succeeded. Return grades and supporting reasons.
\end{quote}
Prompt A checks evidence first. Prompt B checks role requirements and changed content first. Signed changes are computed from these grades.

\label{app:prompts-prediction}
The legacy forward observer instead scores discrepancy directly.
\begin{quote}\small
Compare [reference and perturbed inputs/outputs] using [role and references]. Score four-channel severity and input/output discrepancies in $[0,1]$. Different false content can have equal severity but nonzero discrepancy. Give concise reasons.
\end{quote}
Its historical packet included nominal injection metadata with instructions to ignore it as evidence. This protocol was not fully label-blind and differs from the agreement audit.

\subsection{Injection and Answer Correctness}
\label{app:prompts-injection}
The injector edits content before the executor applies the assigned placement.
\begin{quote}\small
Introduce [error family] into [target text] with a minimal fluent edit. Preserve unrelated content. For factual errors, change a concrete entity, relation, date or quantity. Return the edit and its description.
\end{quote}
A separate judge evaluates the final answer. Task agents do not receive the gold answer.
\begin{quote}\small
Compare [complete output] with [question, reference answer and supporting material]. Accept equivalent answers that satisfy every requested part. Ignore quoted answers that the output rejects. Score abstention as incorrect. Mark indeterminate only when the reference prevents judgment. Give a concise reason.
\end{quote}

\subsection{Disablement and Localization}
\label{app:prompts-localization}
\begin{quote}\small
Disable [role]'s task-processing function for this invocation. Provide no task-specific solution. Emit only [role-compatible unavailable placeholder].
\end{quote}
Healthy uses the original prompt, Style changes only presentation, and restoration reinstates Healthy. The node observer does not receive the active fault prompt, fault labels, or gold answers.
\begin{quote}\small
Given [question, normal role interface, input and output], judge whether a substantive role product is present. Placeholders and generic refusals lack content. Incorrect but substantive products count as present. Return present, missing or indeterminate with a reason.
\end{quote}

\subsection{Evidence Delivery}
\label{app:prompts-recovery}
\begin{quote}\small
Use [frozen packet] as optional evidence for your assigned role. Preserve current claims unless relevant evidence supports correction. Do not invent missing links or change the output interface.
\end{quote}
Placement strategies share this packet and wrapper. REPA-RAG changes delivery position while retaining task prompts and workflow, without receiving correctness feedback.

%% file: appendix/E_limitations_future.tex
\section{Limitations and Future Directions}
\label{app:limitations-future}


Our evaluation covers three Qwen checkpoints and six workflows, leaving generalization to other model families and dynamically changing execution structures open. 
Future work could extend the evaluation to heterogeneous agents and adaptive, tool-using workflows under shared tasks and evaluation protocols. Such comparisons would help distinguish model-dependent responses from recurring structural effects and clarify whether REPA retains its predictive accuracy as agent capabilities and interaction patterns change.

ErrPropBench-10 covers a bounded set of knowledge and reasoning domains, leaving broader domain variation to be explored. Adding domains and controlled experimental settings could reveal further propagation patterns and examine how they relate to foundation-model capabilities, pretraining characteristics and MAS configurations, providing a broader basis for understanding when local responses yield reliable system-level predictions.

%% file: main.bbl
\begin{thebibliography}{51}
\providecommand{\natexlab}[1]{#1}
\providecommand{\url}[1]{\texttt{#1}}
\expandafter\ifx\csname urlstyle\endcsname\relax
  \providecommand{\doi}[1]{doi: #1}\else
  \providecommand{\doi}{doi: \begingroup \urlstyle{rm}\Url}\fi

\bibitem[{Anthropic}(2026)]{anthropic2026opus48}
{Anthropic}.
\newblock Introducing {Claude Opus 4.8}.
\newblock Official model announcement, May 2026.
\newblock URL \url{https://www.anthropic.com/news/claude-opus-4-8}.
\newblock Published May 28, 2026. Accessed September 18, 2026.

\bibitem[Asai et~al.(2024)Asai, Wu, Wang, Sil, and Hajishirzi]{asai2023selfrag}
Akari Asai, Zeqiu Wu, Yizhong Wang, Avirup Sil, and Hannaneh Hajishirzi.
\newblock Self-{RAG}: Learning to retrieve, generate, and critique through
  self-reflection.
\newblock In \emph{The Twelfth International Conference on Learning
  Representations}, 2024.
\newblock URL
  \url{https://proceedings.iclr.cc/paper_files/paper/2024/hash/25f7be9694d7b32d5cc670927b8091e1-Abstract-Conference.html}.

\bibitem[Atherton et~al.(2023)Atherton, Schwartz, Fontana, and
  Hall]{atherton2023glossary}
Daniel Atherton, Reva Schwartz, Peter~C. Fontana, and Patrick Hall.
\newblock The language of trustworthy {AI}: An in-depth glossary of terms.
\newblock NIST Artificial Intelligence 100-3, National Institute of Standards
  and Technology, 2023.
\newblock URL \url{https://nvlpubs.nist.gov/nistpubs/ai/NIST.AI.100-3.pdf}.

\bibitem[Berger et~al.(2016)Berger, Israel, Miller, Parkinson, Reeves, and
  Williams]{berger2016world}
Eugene Berger, George~L. Israel, Charlotte Miller, Brian Parkinson, Andrew
  Reeves, and Nadejda Williams.
\newblock \emph{World History: Cultures, States, and Societies to 1500}.
\newblock University of North Georgia Press, 2016.
\newblock ISBN 9781940771106.
\newblock URL \url{https://oer.galileo.usg.edu/history-textbooks/2/}.

\bibitem[Carbonell and Goldstein(1998)]{carbonell1998mmr}
Jaime Carbonell and Jade Goldstein.
\newblock The use of {MMR}, diversity-based reranking for reordering documents
  and producing summaries.
\newblock In \emph{Proceedings of the 21st Annual International ACM SIGIR
  Conference on Research and Development in Information Retrieval}, pages
  335--336. ACM, 1998.
\newblock \doi{10.1145/290941.291025}.
\newblock URL \url{https://doi.org/10.1145/290941.291025}.

\bibitem[Cardona et~al.(2018)Cardona, Shao, and
  Nixon]{cardona2018photosynthesis}
Tanai Cardona, Shengxi Shao, and Peter~J. Nixon.
\newblock Enhancing photosynthesis in plants: The light reactions.
\newblock \emph{Essays in Biochemistry}, 62\penalty0 (1):\penalty0 85--94,
  2018.
\newblock \doi{10.1042/EBC20170015}.
\newblock URL \url{https://pmc.ncbi.nlm.nih.gov/articles/PMC5897789/}.

\bibitem[Cemri et~al.(2025)Cemri, Pan, Yang, Agrawal, Chopra, Tiwari, Keutzer,
  Parameswaran, Klein, Ramchandran, Zaharia, Gonzalez, and
  Stoica]{cemri2025mast}
Mert Cemri, Melissa~Z. Pan, Shuyi Yang, Lakshya~A. Agrawal, Bhavya Chopra,
  Rishabh Tiwari, Kurt Keutzer, Aditya Parameswaran, Dan Klein, Kannan
  Ramchandran, Matei~A. Zaharia, Joseph~E. Gonzalez, and Ion Stoica.
\newblock Why do multi-agent {LLM} systems fail?
\newblock In \emph{Advances in Neural Information Processing Systems},
  volume~38. Curran Associates, Inc., 2025.
\newblock \doi{10.52202/085713-4082}.
\newblock URL
  \url{https://proceedings.neurips.cc/paper_files/paper/2025/hash/b1041e52d3be19f0a9bc491657488e4a-Abstract-Datasets_and_Benchmarks_Track.html}.
\newblock Datasets and Benchmarks Track.

\bibitem[{DeepSeek-AI}(2026)]{deepseekai2026deepseekv4}
{DeepSeek-AI}.
\newblock {DeepSeek-V4}: Towards highly efficient million-token context
  intelligence, 2026.
\newblock URL \url{https://arxiv.org/abs/2606.19348}.

\bibitem[Emde et~al.(2026{\natexlab{a}})Emde, Goel, Yun, Oh, and
  Gubri]{emde2026lost}
Cornelius Emde, Anmol Goel, Sangdoo Yun, Seong~Joon Oh, and Martin Gubri.
\newblock Lost in communication: Uncertainty propagation in multi-agent
  systems.
\newblock In \emph{ICML 2026 Workshop on Statistical Frameworks for Uncertainty
  in Agentic Systems}, 2026{\natexlab{a}}.
\newblock URL \url{https://gubri.eu/publication/lost_in_communication/}.
\newblock Non-archival workshop paper.

\bibitem[Emde et~al.(2026{\natexlab{b}})Emde, Rubinstein, Goel, Heakl, Yun, Oh,
  and Gubri]{emde2026maseval}
Cornelius Emde, Alexander Rubinstein, Anmol Goel, Ahmed Heakl, Sangdoo Yun,
  Seong~Joon Oh, and Martin Gubri.
\newblock {MASEval}: Extending multi-agent evaluation from models to systems.
\newblock In \emph{Proceedings of the 64th Annual Meeting of the Association
  for Computational Linguistics (Volume 3: System Demonstrations)}, pages
  345--356. Association for Computational Linguistics, 2026{\natexlab{b}}.
\newblock \doi{10.18653/v1/2026.acl-demo.34}.
\newblock URL \url{https://aclanthology.org/2026.acl-demo.34/}.

\bibitem[Ho et~al.(2020)Ho, Duong~Nguyen, Sugawara, and Aizawa]{ho2020twowiki}
Xanh Ho, Anh-Khoa Duong~Nguyen, Saku Sugawara, and Akiko Aizawa.
\newblock Constructing a multi-hop {QA} dataset for comprehensive evaluation of
  reasoning steps.
\newblock In \emph{Proceedings of the 28th International Conference on
  Computational Linguistics}, pages 6609--6625. International Committee on
  Computational Linguistics, 2020.
\newblock \doi{10.18653/v1/2020.coling-main.580}.
\newblock URL \url{https://aclanthology.org/2020.coling-main.580/}.

\bibitem[Hoerl and Kennard(1970)]{hoerl1970ridge}
Arthur~E. Hoerl and Robert~W. Kennard.
\newblock Ridge regression: Biased estimation for nonorthogonal problems.
\newblock \emph{Technometrics}, 12\penalty0 (1):\penalty0 55--67, 1970.
\newblock \doi{10.1080/00401706.1970.10488634}.
\newblock URL \url{https://doi.org/10.1080/00401706.1970.10488634}.

\bibitem[Hong et~al.(2024)Hong, Zhuge, Chen, Zheng, Cheng, Zhang, Wang, Wang,
  Yau, Lin, Zhou, Ran, Xiao, Wu, and Schmidhuber]{hong2024metagpt}
Sirui Hong, Mingchen Zhuge, Jonathan Chen, Xiawu Zheng, Yuheng Cheng, Ceyao
  Zhang, Jinlin Wang, Zili Wang, Steven Ka~Shing Yau, Zijuan Lin, Liyang Zhou,
  Chenyu Ran, Lingfeng Xiao, Chenglin Wu, and J{\"u}rgen Schmidhuber.
\newblock {MetaGPT}: Meta programming for a multi-agent collaborative
  framework.
\newblock In \emph{The Twelfth International Conference on Learning
  Representations}, 2024.
\newblock URL
  \url{https://proceedings.iclr.cc/paper_files/paper/2024/hash/6507b115562bb0a305f1958ccc87355a-Abstract-Conference.html}.

\bibitem[Horn and Johnson(2013)]{horn2012matrix}
Roger~A. Horn and Charles~R. Johnson.
\newblock \emph{Matrix Analysis}.
\newblock Cambridge University Press, second edition, 2013.
\newblock \doi{10.1017/CBO9781139020411}.
\newblock URL \url{https://doi.org/10.1017/CBO9781139020411}.

\bibitem[Huang et~al.(2025)Huang, Zhou, Jin, Zhou, Chen, Wang, Yuan, Lyu, and
  Sap]{huang2025resilience}
Jen-Tse Huang, Jiaxu Zhou, Tailin Jin, Xuhui Zhou, Zixi Chen, Wenxuan Wang,
  Youliang Yuan, Michael Lyu, and Maarten Sap.
\newblock On the resilience of {LLM}-based multi-agent collaboration with
  faulty agents.
\newblock In \emph{Proceedings of the 42nd International Conference on Machine
  Learning}, volume 267 of \emph{Proceedings of Machine Learning Research},
  pages 26202--26226. PMLR, 2025.
\newblock URL \url{https://proceedings.mlr.press/v267/huang25ay.html}.

\bibitem[Jamshidi(2026)]{jamshidi2026collective}
Saeid Jamshidi.
\newblock Collective hallucination in multi-agent {LLMs}: Modeling and defense,
  2026.
\newblock URL \url{https://arxiv.org/abs/2606.07941v2}.
\newblock Preprint, version 2, 18 August 2026.

\bibitem[Jamshidi et~al.(2026)Jamshidi, Moradi~Dakhel, Nafi, and
  Khomh]{jamshidi2026hallucinationcascade}
Saeid Jamshidi, Arghavan Moradi~Dakhel, Kawser~Wazed Nafi, and Foutse Khomh.
\newblock Hallucination cascade: Analyzing error propagation in multi-agent
  {LLM} systems, 2026.
\newblock URL \url{https://arxiv.org/abs/2606.07937v1}.
\newblock Preprint, version 1.

\bibitem[Jia et~al.(2026)Jia, Deng, Chen, Wang, and Zheng]{jia2026masfire}
Jin Jia, Zhiling Deng, Zhuangbin Chen, Yingqi Wang, and Zibin Zheng.
\newblock {MAS-FIRE}: Fault injection and reliability evaluation for
  {LLM}-based multi-agent systems, 2026.
\newblock URL \url{https://arxiv.org/abs/2602.19843}.

\bibitem[Johnson(2016)]{johnson2016photosynthesis}
Matthew~P. Johnson.
\newblock Photosynthesis.
\newblock \emph{Essays in Biochemistry}, 60\penalty0 (3):\penalty0 255--273,
  2016.
\newblock \doi{10.1042/EBC20160016}.
\newblock URL \url{https://pmc.ncbi.nlm.nih.gov/articles/PMC5264509/}.

\bibitem[Kalman(1960)]{kalman1960filtering}
R.~E. Kalman.
\newblock A new approach to linear filtering and prediction problems.
\newblock \emph{Journal of Basic Engineering}, 82\penalty0 (1):\penalty0
  35--45, 1960.
\newblock \doi{10.1115/1.3662552}.
\newblock URL \url{https://doi.org/10.1115/1.3662552}.

\bibitem[Khalil(2002)]{khalil2002nonlinear}
Hassan~K. Khalil.
\newblock \emph{Nonlinear Systems}.
\newblock Prentice Hall, 3 edition, 2002.
\newblock ISBN 0-13-067389-7.
\newblock URL
  \url{https://www.pearson.com/en-ca/subject-catalog/p/nonlinear-systems/P200000003306/9780130675309}.

\bibitem[Lewis et~al.(2020)Lewis, Perez, Piktus, Petroni, Karpukhin, Goyal,
  K{\"u}ttler, Lewis, Yih, Rockt{\"a}schel, Riedel, and
  Kiela]{lewis2020retrieval}
Patrick Lewis, Ethan Perez, Aleksandra Piktus, Fabio Petroni, Vladimir
  Karpukhin, Naman Goyal, Heinrich K{\"u}ttler, Mike Lewis, Wen-tau Yih, Tim
  Rockt{\"a}schel, Sebastian Riedel, and Douwe Kiela.
\newblock Retrieval-augmented generation for knowledge-intensive {NLP} tasks.
\newblock In \emph{Advances in Neural Information Processing Systems},
  volume~33, pages 9459--9474. Curran Associates, Inc., 2020.
\newblock URL
  \url{https://proceedings.neurips.cc/paper_files/paper/2020/hash/6b493230205f780e1bc26945df7481e5-Abstract.html}.

\bibitem[Li et~al.(2023)Li, Hammoud, Itani, Khizbullin, and
  Ghanem]{li2023camel}
Guohao Li, Hasan Abed Al~Kader Hammoud, Hani Itani, Dmitrii Khizbullin, and
  Bernard Ghanem.
\newblock {CAMEL}: Communicative agents for ``mind'' exploration of large
  language model society.
\newblock In \emph{Advances in Neural Information Processing Systems},
  volume~36, 2023.
\newblock URL
  \url{https://proceedings.neurips.cc/paper/2023/hash/a3621ee907def47c1b952ade25c67698-Abstract-Conference.html}.

\bibitem[Lin(1989)]{lin1989concordance}
L.~I. Lin.
\newblock A concordance correlation coefficient to evaluate reproducibility.
\newblock \emph{Biometrics}, 45\penalty0 (1):\penalty0 255--268, 1989.
\newblock \doi{10.2307/2532051}.
\newblock URL \url{https://pubmed.ncbi.nlm.nih.gov/2720055/}.

\bibitem[Liu et~al.(2026)Liu, Huang, and Guo]{liu2026mosaic}
Yiren Liu, Yun Huang, and Shunan Guo.
\newblock {Mosaic}: A multi-level framework for observing and evaluating
  performance of multi-agent systems.
\newblock In \emph{Proceedings of the Extended Abstracts of the 2026 CHI
  Conference on Human Factors in Computing Systems}, CHI EA '26, pages 1--9.
  Association for Computing Machinery, 2026.
\newblock \doi{10.1145/3772363.3798830}.
\newblock URL \url{https://doi.org/10.1145/3772363.3798830}.

\bibitem[Liu(2026)]{liu2026contagion}
Zewen Liu.
\newblock Contagion networks: Evaluator preference propagation in multi-agent
  {LLM} systems, 2026.
\newblock URL \url{https://arxiv.org/abs/2606.20493v2}.
\newblock Preprint, version 2, 26 June 2026.

\bibitem[Madaan et~al.(2023)Madaan, Tandon, Gupta, Hallinan, Gao, Wiegreffe,
  Alon, Dziri, Prabhumoye, Yang, Gupta, Majumder, Hermann, Welleck,
  Yazdanbakhsh, and Clark]{madaan2023selfrefine}
Aman Madaan, Niket Tandon, Prakhar Gupta, Skyler Hallinan, Luyu Gao, Sarah
  Wiegreffe, Uri Alon, Nouha Dziri, Shrimai Prabhumoye, Yiming Yang, Shashank
  Gupta, Bodhisattwa~Prasad Majumder, Katherine Hermann, Sean Welleck, Amir
  Yazdanbakhsh, and Peter Clark.
\newblock Self-refine: Iterative refinement with self-feedback.
\newblock In \emph{Advances in Neural Information Processing Systems},
  volume~36, pages 46534--46594. Curran Associates, Inc., 2023.
\newblock \doi{10.52202/075280-2019}.
\newblock URL
  \url{https://proceedings.neurips.cc/paper_files/paper/2023/hash/91edff07232fb1b55a505a9e9f6c0ff3-Abstract-Conference.html}.

\bibitem[{National Human Genome Research Institute}(2020)]{nhgri2020dna}
{National Human Genome Research Institute}.
\newblock Deoxyribonucleic acid ({DNA}) fact sheet.
\newblock National Institutes of Health, 2020.
\newblock URL
  \url{https://www.genome.gov/about-genomics/fact-sheets/Deoxyribonucleic-Acid-Fact-Sheet}.
\newblock Updated August 24, 2020.

\bibitem[Niu et~al.(2026)Niu, Shu, and Zhao]{niu2026reliability}
Ruiwu Niu, Xincheng Shu, and Ying Zhao.
\newblock Reliability--contagion feasibility in {LLM} multi-agent networks,
  2026.
\newblock URL \url{https://arxiv.org/abs/2607.21912v1}.
\newblock Preprint, version 1, 24 July 2026.

\bibitem[Nocedal and Wright(2006)]{nocedal2006numerical}
Jorge Nocedal and Stephen~J. Wright.
\newblock \emph{Numerical Optimization}.
\newblock Springer, 2 edition, 2006.
\newblock \doi{10.1007/978-0-387-40065-5}.
\newblock URL \url{https://doi.org/10.1007/978-0-387-40065-5}.

\bibitem[{OpenAI}(2026)]{openai2026gpt56}
{OpenAI}.
\newblock {GPT-5.6}: Frontier intelligence that scales with your ambition.
\newblock Official model announcement, July 2026.
\newblock URL \url{https://openai.com/index/gpt-5-6/}.
\newblock Published July 9, 2026. Accessed September 18, 2026.

\bibitem[Pan et~al.(2026)Pan, Yu, Wu, Ye, Si, Xu, and Wu]{pan2026sfcot}
Yu~Pan, Wenlong Yu, Tiejun Wu, Xiaohu Ye, Qiannan Si, Guangquan Xu, and Bin Wu.
\newblock {SFCoT}: Safer chain-of-thought via active safety evaluation and
  calibration.
\newblock In \emph{ICASSP 2026 - 2026 IEEE International Conference on
  Acoustics, Speech and Signal Processing (ICASSP)}, pages 19837--19841. IEEE,
  2026.
\newblock \doi{10.1109/ICASSP55912.2026.11463331}.
\newblock URL \url{https://doi.org/10.1109/ICASSP55912.2026.11463331}.

\bibitem[{Qwen Team}(2024)]{qwen2024qwen25}
{Qwen Team}.
\newblock {Qwen2.5-14B-Instruct}: Official model card, 2024.
\newblock URL \url{https://huggingface.co/Qwen/Qwen2.5-14B-Instruct}.

\bibitem[{Qwen Team}(2026{\natexlab{a}})]{qwen2026qwen35}
{Qwen Team}.
\newblock {Qwen3.5-9B}: Official model card, 2026{\natexlab{a}}.
\newblock URL \url{https://huggingface.co/Qwen/Qwen3.5-9B}.

\bibitem[{Qwen Team}(2026{\natexlab{b}})]{qwen2026qwen35moe}
{Qwen Team}.
\newblock {Qwen3.5-122B-A10B}: Official model card, 2026{\natexlab{b}}.
\newblock URL \url{https://huggingface.co/Qwen/Qwen3.5-122B-A10B}.

\bibitem[Reddy(2020)]{reddy2020blackholes}
Francis Reddy.
\newblock What are black holes?
\newblock NASA, 2020.
\newblock URL \url{https://www.nasa.gov/universe/what-are-black-holes/}.
\newblock Published September 8, 2020; updated July 26, 2023.

\bibitem[Robertson and Zaragoza(2009)]{robertson2009bm25}
Stephen~E. Robertson and Hugo Zaragoza.
\newblock The probabilistic relevance framework: {BM25} and beyond.
\newblock \emph{Foundations and Trends in Information Retrieval}, 3\penalty0
  (4):\penalty0 333--389, 2009.
\newblock \doi{10.1561/1500000019}.
\newblock URL \url{https://doi.org/10.1561/1500000019}.

\bibitem[Shen et~al.(2025)Shen, Liu, Dai, Wang, Miao, Tan, Pan, and
  Wang]{shen2025eib}
Xu~Shen, Yixin Liu, Yiwei Dai, Yili Wang, Rui Miao, Yue Tan, Shirui Pan, and
  Xin Wang.
\newblock Understanding the information propagation effects of communication
  topologies in {LLM}-based multi-agent systems.
\newblock In \emph{Proceedings of the 2025 Conference on Empirical Methods in
  Natural Language Processing}, pages 12347--12361, Suzhou, China, November
  2025. Association for Computational Linguistics.
\newblock \doi{10.18653/v1/2025.emnlp-main.623}.
\newblock URL \url{https://aclanthology.org/2025.emnlp-main.623/}.

\bibitem[Shen et~al.(2026)Shen, Zhang, Wang, Tan, Zhao, Yao, Tadiparthi,
  Mahjoub, Pari, Lee, and Chen]{shen2026masc}
Xu~Shen, Qi~Zhang, Song Wang, Zhen Tan, Xinyu Zhao, Laura Yao, Vaishnav
  Tadiparthi, Hossein~Nourkhiz Mahjoub, Ehsan~Moradi Pari, Kwonjoon Lee, and
  Tianlong Chen.
\newblock Metacognitive self-correction for multi-agent system via
  prototype-guided next-execution reconstruction.
\newblock In \emph{Findings of the Association for Computational Linguistics:
  ACL 2026}, pages 23320--23337, San Diego, California, USA, 2026. Association
  for Computational Linguistics.
\newblock URL \url{https://aclanthology.org/2026.findings-acl.1168.pdf}.

\bibitem[Shinn et~al.(2023)Shinn, Cassano, Gopinath, Narasimhan, and
  Yao]{shinn2023reflexion}
Noah Shinn, Federico Cassano, Ashwin Gopinath, Karthik Narasimhan, and Shunyu
  Yao.
\newblock Reflexion: Language agents with verbal reinforcement learning.
\newblock In \emph{Advances in Neural Information Processing Systems},
  volume~36, pages 8634--8652. Curran Associates, Inc., 2023.
\newblock \doi{10.52202/075280-0377}.
\newblock URL
  \url{https://proceedings.neurips.cc/paper_files/paper/2023/hash/1b44b878bb782e6954cd888628510e90-Abstract-Conference.html}.

\bibitem[Trivedi et~al.(2022)Trivedi, Balasubramanian, Khot, and
  Sabharwal]{trivedi2022musique}
Harsh Trivedi, Niranjan Balasubramanian, Tushar Khot, and Ashish Sabharwal.
\newblock {MuSiQue}: Multihop questions via single-hop question composition.
\newblock \emph{Transactions of the Association for Computational Linguistics},
  10:\penalty0 539--554, 2022.
\newblock \doi{10.1162/tacl_a_00475}.
\newblock URL \url{https://aclanthology.org/2022.tacl-1.31/}.

\bibitem[Wu et~al.(2023)Wu, Bansal, Zhang, Wu, Li, Zhu, Jiang, Zhang, Zhang,
  Liu, Awadallah, White, Burger, and Wang]{wu2023autogen}
Qingyun Wu, Gagan Bansal, Jieyu Zhang, Yiran Wu, Beibin Li, Erkang Zhu,
  Li~Jiang, Xiaoyun Zhang, Shaokun Zhang, Jiale Liu, Ahmed~Hassan Awadallah,
  Ryen~W. White, Doug Burger, and Chi Wang.
\newblock {AutoGen}: Enabling next-gen {LLM} applications via multi-agent
  conversation, 2023.
\newblock URL \url{https://arxiv.org/abs/2308.08155}.

\bibitem[Yaga et~al.(2018)Yaga, Mell, Roby, and Scarfone]{yaga2018blockchain}
Dylan Yaga, Peter Mell, Nik Roby, and Karen Scarfone.
\newblock Blockchain technology overview.
\newblock NIST Interagency Report 8202, National Institute of Standards and
  Technology, 2018.
\newblock URL \url{https://csrc.nist.gov/pubs/ir/8202/final}.

\bibitem[Yang et~al.(2018)Yang, Qi, Zhang, Bengio, Cohen, Salakhutdinov, and
  Manning]{yang2018hotpotqa}
Zhilin Yang, Peng Qi, Saizheng Zhang, Yoshua Bengio, William Cohen, Ruslan
  Salakhutdinov, and Christopher~D. Manning.
\newblock {HotpotQA}: A dataset for diverse, explainable multi-hop question
  answering.
\newblock In \emph{Proceedings of the 2018 Conference on Empirical Methods in
  Natural Language Processing}, pages 2369--2380. Association for Computational
  Linguistics, 2018.
\newblock \doi{10.18653/v1/D18-1259}.
\newblock URL \url{https://aclanthology.org/D18-1259/}.

\bibitem[Yu et~al.(2025)Yu, Wang, Zhang, Mao, Yin, Liu, Wang, Wen, and
  Wang]{yu2025netsafe}
Miao Yu, Shilong Wang, Guibin Zhang, Junyuan Mao, Chenlong Yin, Qijiong Liu,
  Kun Wang, Qingsong Wen, and Yang Wang.
\newblock {NetSafe}: Exploring the topological safety of multi-agent system.
\newblock In \emph{Findings of the Association for Computational Linguistics:
  {ACL} 2025}, pages 2905--2938. Association for Computational Linguistics,
  2025.
\newblock \doi{10.18653/v1/2025.findings-acl.150}.
\newblock URL \url{https://aclanthology.org/2025.findings-acl.150/}.

\bibitem[{Z.ai}(2026)]{zai2026glm52}
{Z.ai}.
\newblock {GLM-5.2}.
\newblock Hugging Face model card, 2026.
\newblock URL \url{https://huggingface.co/zai-org/GLM-5.2}.
\newblock MIT license. Accessed September 18, 2026.

\bibitem[Zhang et~al.(2023)Zhang, Lipton, Li, and Smola]{zhang2023dive}
Aston Zhang, Zachary~C. Lipton, Mu~Li, and Alexander~J. Smola.
\newblock \emph{Dive into Deep Learning}.
\newblock Cambridge University Press, 2023.
\newblock URL \url{https://d2l.ai/}.

\bibitem[Zhang et~al.(2025)Zhang, Yin, Zhang, Liu, Han, Zhang, Li, Wang, Wang,
  Chen, and Wu]{zhang2025whowhen}
Shaokun Zhang, Ming Yin, Jieyu Zhang, Jiale Liu, Zhiguang Han, Jingyang Zhang,
  Beibin Li, Chi Wang, Huazheng Wang, Yiran Chen, and Qingyun Wu.
\newblock Which agent causes task failures and when? {On} automated failure
  attribution of {LLM} multi-agent systems.
\newblock In \emph{Proceedings of the 42nd International Conference on Machine
  Learning}, volume 267 of \emph{Proceedings of Machine Learning Research},
  pages 76583--76599. PMLR, 2025.
\newblock URL \url{https://proceedings.mlr.press/v267/zhang25cq.html}.

\bibitem[Zheng et~al.(2023)Zheng, Chiang, Sheng, Zhuang, Wu, Zhuang, Lin, Li,
  Li, Xing, Zhang, Gonzalez, and Stoica]{zheng2023judging}
Lianmin Zheng, Wei-Lin Chiang, Ying Sheng, Siyuan Zhuang, Zhanghao Wu, Yonghao
  Zhuang, Zi~Lin, Zhuohan Li, Dacheng Li, Eric~P. Xing, Hao Zhang, Joseph~E.
  Gonzalez, and Ion Stoica.
\newblock Judging {LLM-as-a-Judge} with {MT-Bench} and {Chatbot Arena}.
\newblock In \emph{Advances in Neural Information Processing Systems},
  volume~36, pages 46595--46623, 2023.
\newblock \doi{10.52202/075280-2020}.
\newblock URL
  \url{https://proceedings.neurips.cc/paper_files/paper/2023/hash/91f18a1287b398d378ef22505bf41832-Abstract-Datasets_and_Benchmarks.html}.

\bibitem[Zhou and Yao(2023)]{zhou2023crispr}
Lifang Zhou and Shaohua Yao.
\newblock Recent advances in therapeutic {CRISPR-Cas9} genome editing:
  Mechanisms and applications.
\newblock \emph{Molecular Biomedicine}, 4\penalty0 (1):\penalty0 10, 2023.
\newblock \doi{10.1186/s43556-023-00115-5}.
\newblock URL
  \url{https://link.springer.com/article/10.1186/s43556-023-00115-5}.

\bibitem[Zhou et~al.(2026)Zhou, Liu, Liu, Shao, Wang, Shao, Jin, and
  Xu]{zhou2026resmas}
Zhilun Zhou, Zihan Liu, Jiahe Liu, Qingyu Shao, Yihan Wang, Kun Shao, Depeng
  Jin, and Fengli Xu.
\newblock {ResMAS}: Resilience optimization in {LLM}-based multi-agent systems.
\newblock In \emph{Proceedings of the AAAI Conference on Artificial
  Intelligence}, volume~40, pages 35176--35184. AAAI Press, 2026.
\newblock \doi{10.1609/aaai.v40i41.40824}.
\newblock URL \url{https://ojs.aaai.org/index.php/AAAI/article/view/40824}.

\end{thebibliography}
